\documentclass[preprint,review,12pt]{elsarticle}

\biboptions{numbers,sort&compress}

\usepackage[english]{babel}
\usepackage{microtype}
\usepackage{placeins}
\usepackage{flafter}
\usepackage{subcaption}
\usepackage{booktabs}
\usepackage{amssymb}
\usepackage{mathtools}
\usepackage{algorithm}
\usepackage{algpseudocode}
\usepackage{siunitx}
\usepackage{tikz}
\usepackage{csquotes}
\usepackage{url}
\usepackage{hyperref}
\usetikzlibrary{arrows.meta,positioning,fit,calc}

\DeclarePairedDelimiter{\norm}{\lVert}{\rVert}

\makeatletter
\def\ps@pprintTitle{%
 \let\@oddhead\@empty
 \let\@evenhead\@empty
 \def\@oddfoot{\centerline{\thepage}}%
 \let\@evenfoot\@oddfoot}
\makeatother

\begin{document}

\begin{frontmatter}

\title{Tensor-Based Modal Decomposition and Sparse Sensor Placement for the Brugge Field Simulation Model}

\author[inst1]{D. Samatov\corref{cor1}\fnref{orcid1}}
\ead{dss53@tpu.ru}
\fntext[orcid1]{ORCID: 0009-0000-1821-323X}

\author[inst1]{B. Merzlikin\fnref{orcid2}}
\fntext[orcid2]{ORCID: 0000-0001-8545-9491}

\author[inst1]{G. Shishaev\fnref{orcid3}}
\fntext[orcid3]{ORCID: 0000-0003-2387-9217}

\affiliation[inst1]{organization={National Research Tomsk Polytechnic University},
                    city={Tomsk},
                    postcode={634050},
                    country={Russia}}

\cortext[cor1]{Corresponding author.}

\begin{abstract}
Sparse monitoring of reservoir pressure and saturation requires numerical methods that retain gridded structure while reconstructing full fields from few observations. We present a four-dimensional tensor-based modal decomposition (TBMD) and sparse reconstruction framework for coupled pressure--saturation fields. The approach uses an explicit property mode, mode-4 pivoted orthogonal--triangular (QR) ranking of grid-wide spatial--property fibers, and tensor-based compressive sensing for both grid-wide and existing-well measurement operators. The Brugge benchmark is evaluated using 10 well-control realizations with fixed geology. Each realization is processed independently as $\mathcal{X}^{(r)}\in\mathbb{R}^{139\times48\times2\times134}$ under an 80/20 temporal split with training-fitted property-wise min--max normalization. In the joint pressure--saturation well-only study, increasing the number of instrumented wells from 1 to 10 reduces the relative Frobenius error from about 0.57 to 0.20, increases the Structural Similarity Index Measure from about 0.47 to 0.88, and raises the peak signal-to-noise ratio from about 33.8 to 37 dB; by 20 wells, the relative error drops to about 0.11. The results support the feasibility of tensor-structured sparse reconstruction for coupled reservoir fields and provide a basis for controlled comparisons with alternative reduced-order and sparse-sensing methods.
\end{abstract}

\begin{highlights}
\item Four-dimensional TBMD couples pressure, saturation, space, and time.
\item Mode-4 pivoted QR ranks grid-wide spatial--property sensor fibers.
\item Tensor compressive sensing reconstructs held-out field snapshots.
\item Well-only tests separate operational constraints from grid-wide QR.
\item Held-out Brugge tests evaluate sparse full-field reconstruction.
\end{highlights}

\begin{keyword}
Tensor-based modal decomposition \sep Sparse sensor placement \sep Tensor compressive sensing \sep Reservoir simulation \sep Reduced-order modeling \sep Scientific machine learning
\end{keyword}

\end{frontmatter}

\section{Introduction}
Modern reservoir management increasingly seeks methods that could support more timely monitoring of pressure and fluid saturation. Such vigilance is fundamental to optimizing well operations and waterflooding strategies. However, acquiring direct measurements of reservoir pressure constitutes a complex challenge. Specifically, large-scale pressure surveys remain costly and often require production shutdown for well testing. Moreover, selecting informative observation points is a non-trivial task, resulting in limited reservoir state information and diminished management efficiency~\cite{manohar2018}.

Existing numerical approaches include history matching, matrix reduced-order models, tensor decompositions, sparse sensor selection, compressive sensing, and deep learning. Section~\ref{sec:related_work} positions the proposed tensor workflow relative to these methods and clarifies how the present manuscript uses previous baseline evidence.

Building upon the foundational work of Zhong et al.~\cite{zhong2024tbmd}, we generalize TBMD to a four-dimensional formulation by introducing an explicit property mode. This enables the simultaneous and consistent processing of multiple physical quantities, such as pressure and oil saturation. 

\textbf{Contributions.} The main contributions of this paper are:
\begin{enumerate}
  \item A fourth-order tensor formulation that jointly represents pressure and saturation with an explicit property mode, preserving the gridded spatial organization of the reservoir state.
  \item A mode-4 fiber-pivoted QR ranking algorithm for grid-wide spatial--property sensor placement on the tensor dictionary.
  \item A unified TBCS reconstruction solver evaluated under both grid-wide and existing-well measurement constraints.
  \item A fixed-geology held-out temporal validation on the Brugge benchmark, quantifying the trade-off between monitoring accuracy and the number of measurements.
\end{enumerate}

We validate our proposed workflow on the Brugge field benchmark case using 10 control realizations. Each run is processed independently as a fourth-order tensor, applying an 80/20 temporal split. Property-wise min--max normalization parameters are fitted on that run's training split only.

All pressure values are reported in bars (\si{bar}) when physical units are shown, where $1~\mathrm{bar}=14.5038~\mathrm{psi}$. Reconstruction accuracy is quantified on a common normalized scale via three complementary performance metrics: the relative Frobenius norm error, the Structural Similarity Index Measure (SSIM) with a Gaussian window of $11\times11$ ($\sigma=1.5$), and the Peak Signal-to-Noise Ratio (PSNR) with $\mathrm{MAX}=1$.

\par

\section{Related Work}
\label{sec:related_work}
Traditional history matching---the conventional approach to parameter estimation and data assimilation---is computationally intensive, prone to non-uniqueness, and ill-suited for time-constrained operational decision-making~\cite{oliver2008,jansen2011,evensen2009}. Data-driven reduced-order modeling therefore offers an attractive alternative when many flow snapshots are available. Matrix-based approaches, including Proper Orthogonal Decomposition (POD) and Dynamic Mode Decomposition (DMD), are standard tools for extracting coherent structures from high-dimensional flow data~\cite{berkooz1993,schmid2010,kutz2016dmd}. In the specific context of reservoir simulation, POD and its extensions are frequently used for subsurface flow simulation and waterflooding optimization~\cite{vandoren2006,cardoso2010,he2014}. Their limitation in the present setting is that gridded pressure--saturation data must be reshaped into matrices, which obscures the native spatial and property-mode organization of the reservoir state.

Tensor methods provide a natural representation for multidimensional simulation outputs. Higher-order singular value decomposition and multilinear tensor decompositions preserve mode-specific structure and are widely used for compression and reduced-order modeling~\cite{deLathauwer2000,kolda2009}. Tensor extensions of dynamic-mode ideas have also been developed for structured dynamical data~\cite{klus2018}. Tensor-based modal decomposition (TBMD) was recently proposed for reduced-order modeling and sparse sensor placement in a third-order setting~\cite{zhong2024tbmd}. The present work extends this direction by introducing an explicit fourth property mode for joint pressure--saturation modeling and by separating grid-wide QR placement from well-constrained measurement scenarios.

Sparse sensing and reconstruction are closely related to Discrete Empirical Interpolation Method (DEIM), Q-DEIM, pivoted QR, and determinant-based greedy selection~\cite{chaturantabut2010,drmac2016_qdeim,saito2021}. These approaches select informative rows or measurement locations for matrix bases. Our mode-4 fiber-pivoted QR adapts this principle to a tensor dictionary whose columns correspond to spatial--property fibers. The reconstruction stage is based on compressive sensing and sparse recovery~\cite{donoho2006,candes2007}, including multidimensional and block-sparse variants~\cite{caiafa2013multidimcs,eldar2010_block}. Deep-learning field reconstruction from sparse sensors and surrogate modeling under time-varying well controls are other relevant directions~\cite{fukami2021_vm,jin2020}; however, the tensor workflow used here is deliberately linear-algebraic and interpretable, which is useful for diagnosing sensor placement behavior. The core TBMD methodology was previously evaluated by Zhong et al.~\cite{zhong2024tbmd} against POD and DMD for modal decomposition. The present manuscript focuses on its extension to the coupled Brugge reservoir setting.

\section{Mathematical Formulation}
\label{sec:methodology}
\par
The proposed methodology~\cite{zhong2024tbmd} integrates three tensor components into a unified workflow for reduced-order modeling and dynamic reconstruction of the reservoir states. The workflow reuses the core elements of the TBMD framework evaluated by Zhong et al.~\cite{zhong2024tbmd}, while the new formulation in this manuscript is the reservoir-specific fourth-order pressure--saturation construction and its associated grid-wide and existing-well measurement operators.

First, we apply TBMD directly to multidimensional reservoir data and extract dominant spatiotemporal modes while retaining the gridded spatial organization and cross-property structure of the simulated fields. In contrast to ~\cite{zhong2024tbmd}, we generalize TBMD to a fourth-order tensor representation. Each control realization $r\in\{1,\dots,10\}$---that is, each realization of operational controls under fixed geology---is represented by its own tensor
\[
\mathcal{X}^{(r)}\in\mathbb{R}^{I\times J\times K\times L},
\]
where $I$, $J$, and $L$ denote the spatial ($x,y$) and temporal ($t$) modes, respectively. In the joint setting considered throughout the manuscript, the property mode has size $K=2$ and follows the fixed convention stated in Sec.~\ref{subsec:conventions_units}. This ordering is used consistently in the stacked measurement operators, sensing matrices, and training-fitted property-wise min--max normalization. The run index is treated as an external case label rather than as a fifth tensor mode: TBMD, grid-wide QR, and TBCS are executed independently for each run, and the resulting held-out metrics are aggregated only after the run-wise evaluations are complete. For notational simplicity, the superscript $(r)$ is omitted in the generic formulas below whenever a statement applies to every run separately. This extension represents coupled pressure and oil-saturation fields in one tensor dictionary for subsequent sensor ranking and field reconstruction. Then, a tensor QR factorization with mode-4 pivoting ranks spatial--property fibers according to the dictionary residual criterion, thereby defining QR-guided grid-wide sensor placements under budgetary constraints. Finally, TBCS is applied at the snapshot level: for each held-out time step it recovers the corresponding 3D field from sparse scalar observations via an $\ell_1$-regularized regression solved using ADMM, and repeating this recovery over time yields the reconstructed temporal trajectory. In parallel, the same reconstruction stage is used to assess well-constrained measurements at existing wells, and the grid-wide QR selections are later mapped to the well network for operational interpretation.
\par
The workflow component details and corresponding algorithms are provided in the following subsections.

\subsection{Problem Statement}
\label{subsec:problem_statement}
\par
For each control realization $r$, the available training data are the first 80\% of the property-wise normalized reservoir tensor, denoted $\widetilde{\mathcal{X}}^{(r)}_{\mathrm{train}}\in\mathbb{R}^{139\times48\times2\times107}$. The held-out data are the remaining 27 snapshots, each represented as a normalized pressure--saturation field $\mathcal{Z}^{(r,\ell)}\in\mathbb{R}^{139\times48\times2}$ with pressure at property index $k=1$ and oil saturation at $k=2$. The goal is to fit a run-wise TBMD dictionary $\mathcal{A}^{(r)}\in\mathbb{R}^{I\times J\times K\times W}$ from the training tensor and reconstruct each held-out snapshot from a limited set of measurements.
\par
Two admissible measurement regimes are considered. In the grid-wide regime, measurements are scalar spatial--property samples from $\Omega_{\mathrm{grid}}\subseteq\{1,\ldots,I\}\times\{1,\ldots,J\}\times\{1,\ldots,K\}$, and the active set $\Omega_N$ is obtained from the QR-guided residual-criterion ranking. In the existing-well regime, measurements are restricted to subsets $\mathcal{W}_N\subseteq\mathcal{W}$ of the 30 physical wells, with selected wells mapped to their parent grid cells and stacked by property when both pressure and saturation are used. Given the corresponding measurement operator $\mathcal{M}_N$ and measurement vector $\mathbf{y}_N$, the snapshot-level reconstruction problem is to estimate modal coefficients by the relaxed sparse objective in Eq.~\eqref{eq:l1_l2_problem} and synthesize $\hat{\mathcal{Z}}=\mathcal{A}\times_{(4)}\hat{\mathbf{x}}$. Evaluation uses held-out relative Frobenius error, SSIM with an $11\times11$ Gaussian window ($\sigma=1.5$), and PSNR with $\mathrm{MAX}=1$ on the same training-fitted normalized scale.

\subsection{Conventions and Units}
\label{subsec:conventions_units}
\par
We adopt the fixed mode order
\[
(1)\,x,\qquad (2)\,y,\qquad (3)\,\text{property},\qquad (4)\,\text{time}.
\]
Throughout the manuscript, the property index follows the convention $k=1$ for pressure and $k=2$ for oil saturation. Let $W$ denote the dictionary depth, i.e., the number of retained basis tensors (dictionary atoms) used in the downstream QR and TBCS stages, and let $N$ denote the sensor budget. We explicitly distinguish between a persistent sensor location, a per-snapshot scalar measurement, and the associated temporal fiber. A grid-wide sensor location is a fixed spatial--property index
\[
q=(i_q,j_q,k_q)\in\Omega_{\mathrm{grid}},
\qquad
\Omega_{\mathrm{grid}}\subseteq\{1,\dots,I\}\times\{1,\dots,J\}\times\{1,\dots,K\}.
\]
At a given time step $\ell$, this location returns one scalar measurement
\[
y_q^{(\ell)}=\mathcal{X}_{i_q j_q k_q \ell}.
\]
The full temporal fiber $\mathcal{X}_{i_q j_q k_q:}$ is therefore the sequence $\{y_q^{(\ell)}\}_{\ell=1}^{L}$ of scalar measurements over time; it is not the snapshot-level observation vector passed directly to the TBCS solver. For the well-constrained experiments, we use a distinct symbol for the physical well set,
\[
\mathcal{W}=\{w_1,\dots,w_{30}\},
\qquad
|\mathcal{W}|=30,
\]
where each well $w\in\mathcal{W}$ is mapped to its parent grid cell $(i_w,j_w)$.
Accordingly, the budget unit depends on the sensing regime. In the \emph{grid-wide} experiments, $N$ counts selected spatial--property locations from $\Omega_{\mathrm{grid}}$. In the \emph{well-only} experiments, $N$ counts selected physical wells from $\mathcal{W}$, not individual well--property channels. Thus, even in the joint pressure+saturation setting, the well-only budget is capped at 30 physical wells rather than $30\times2$ property-specific channels.
The manuscript therefore distinguishes three sensing setups: (i) direct grid-wide QR placement on $\Omega_{\mathrm{grid}}$; (ii) well-only reconstruction tests in which the active measurement set is restricted to subsets $\mathcal{W}_N\subseteq\mathcal{W}$; and (iii) a post hoc QR-to-well mapping that translates grid-wide QR hits into well-level significance scores. In the present manuscript, case~(ii) is implemented as a random-subset feasibility/stress test and does \emph{not} include a separate QR reranking over a well-restricted admissible set.
\par
All experiments use a consistent 80/20 temporal split within each simulation run: the first 80\% of time steps form the training segment used to fit TBMD and all downstream components, while the remaining 20\% form a strictly held-out test segment. Unless stated otherwise, all reported metrics are computed on this held-out segment. Summary curves in Sec.~\ref{sec:results} first compute the relevant metric on each run separately and then aggregate the resulting held-out values over the 10 well-control scenarios; in the well-only study, this aggregation additionally spans the random well subsets generated for each budget $N$.
For each run $r$ and property index $k\in\{1,2\}$, let $\mathcal{T}_{\mathrm{train}}$ denote the training time-index set. The min--max normalization parameters are then fitted on the training split only,
\[
a_{\min}^{(r,k)} \,=\, \min_{i,j,\,\ell\in\mathcal{T}_{\mathrm{train}}} \mathcal{X}^{(r)}_{ijk\ell},
\qquad
a_{\max}^{(r,k)} \,=\, \max_{i,j,\,\ell\in\mathcal{T}_{\mathrm{train}}} \mathcal{X}^{(r)}_{ijk\ell},
\]
and define the normalized tensor
\[
\widetilde{\mathcal{X}}^{(r)}_{ijk\ell}
\,=\,
\frac{\mathcal{X}^{(r)}_{ijk\ell}-a_{\min}^{(r,k)}}{a_{\max}^{(r,k)}-a_{\min}^{(r,k)}}.
\]
Thus, the training segment of each property is mapped to $[0,1]$, and the same affine transform is applied unchanged to the held-out test segment. Unless stated otherwise, TBMD basis construction, tensor-QR ranking, measurement extraction, TBCS recovery, and the reported relative Frobenius error, SSIM, and PSNR are all computed on $\widetilde{\mathcal{X}}^{(r)}$; native units are restored only when a reconstructed field is converted back for visualization or physical interpretation (e.g., pressure in bars).

\subsection{Tensor-Based Modal Decomposition}
\label{subsec:tbmd}
\par
As was mentioned above, TBMD is a method for extracting low-dimensional modes from high-dimensional dynamical systems while preserving the spatio-temporal structure of the data. In canonical form, TBMD was applied to third-order tensors (two spatial dimensions and time); here we generalize it to the fourth order by adding a mode that represents multiple physical reservoir properties (e.g., pressure and saturation). We implement TBMD via HOSVD and write it as
\begin{equation}
\mathcal{X} \;=\; \mathcal{G}\;\times_1 U^{(1)}\;\times_2 U^{(2)}\;\times_3 U^{(3)}\;\times_4 U^{(4)} \;+\; \mathcal{E},
\label{eq:tbmd_decomposition_4d}
\end{equation}
where $\mathcal{X}\in\mathbb{R}^{I\times J\times K\times L}$ is the observation tensor, $\mathcal{G}$ is the core, $U^{(n)}$ are the mode factors, and $\mathcal{E}$ is the residual. In the truncated HOSVD implementation used here, the mode factors are obtained from the dominant left singular vectors of the mode unfoldings, and the truncated reconstruction is chosen to target a small Frobenius-norm error,
\begin{equation}
\big\|\mathcal{X} - \hat{\mathcal{X}}\big(U^{(1)},U^{(2)},U^{(3)},U^{(4)},\mathcal{G}\big)\big\|_{F},
\label{eq:tbmd_optimization_4d}
\end{equation}
which we use as a target criterion for truncation quality rather than as a globally solved optimization problem. Accordingly, the retained multilinear ranks are selected to meet prescribed error/energy thresholds, not as a consequence of an exact best-approximation claim for HOSVD. Equations~\eqref{eq:tbmd_decomposition_4d}--\eqref{eq:tbmd_optimization_4d} define the truncated tensor representation and its fitting criterion.
Elementwise,
\begin{equation}
\mathcal{X}_{ijkl} \;=\;
\sum_{p=1}^{P}\sum_{q=1}^{Q}\sum_{r=1}^{R}\sum_{s=1}^{S}
\mathcal{G}_{pqrs}\; U^{(1)}_{ip}\, U^{(2)}_{jq}\, U^{(3)}_{kr}\, U^{(4)}_{ls}
\;+\; \mathcal{E}_{ijkl},
\label{eq:tbmd_elementwise_4d}
\end{equation}
for $i=1,\dots,I$, $j=1,\dots,J$, $k=1,\dots,K$, $l=1,\dots,L$, and $p=1,\dots,P$, $q=1,\dots,Q$, $r=1,\dots,R$, $s=1,\dots,S$. Equation~\eqref{eq:tbmd_elementwise_4d} gives the same decomposition in elementwise form. This representation yields modes with a spatial footprint, temporal evolution, and an additional parametric dimension, enabling a compact description of the coupled dynamics of multiple physical quantities. 
\par
In what follows, $\times_n$ denotes the mode-$n$ product; $\mathcal{X}_{(4)}$ is the mode-4 unfolding (matricization); and $\times_{(4)}\mathbf{x}$ denotes contraction of a dictionary tensor with a coefficient vector. The dictionary $\mathcal{A}\in\mathbb{R}^{I\times J\times K\times W}$ is constructed by stacking $W$ retained basis tensors $\{\mathcal{A}_w\}_{w=1}^W$; here $W$ is the dictionary depth used by the downstream QR and TBCS stages. Consequently, the fourth mode of $\mathcal{A}$ indexes retained dictionary atoms rather than physical time. We denote by $\mathcal{A}_{(4)}\in\mathbb{R}^{W\times(IJK)}$ the corresponding mode-4 unfolding; its columns are in one-to-one correspondence with spatial--property indices $(i,j,k)$. The dictionary is then passed to the QR factorization (Sec.~\ref{subsec:tensor_qr}) and reconstruction problem (Sec.~\ref{subsec:tensor_cs}).

\subsection{Tensor QR Factorization for QR-Guided Sensor Ranking}
\label{subsec:tensor_qr}

\par
To select informative sensor locations within the TBMD framework, we employ a mode-4 fiber–pivot tensor QR—a strategy analogous to pivoted QR used for point selection in matrix reduced-order models (e.g., DEIM~\cite{chaturantabut2010} and Q-DEIM~\cite{drmac2016_qdeim}), now extended to operate directly on the tensor dictionary~\cite{saito2021}.
The original data tensor follows the mode order
\((1)\,x,\ (2)\,y,\ (3)\,\text{property},\ (4)\,\text{time}\).
After truncation, however, the dictionary $\mathcal{A}$ stacks $W$ retained basis tensors along its fourth mode, so $W$ is again the dictionary depth. Hence a dictionary fiber $\mathcal{A}_{ijk:}$ is the vector of modal amplitudes associated with spatial--property index $(i,j,k)$, not a physical time series. Let $M_\Omega \subseteq \{1,\dots,I\}\times\{1,\dots,J\}\times\{1,\dots,K\}$ denote a generic spatial--property availability mask and
\[
\Omega = \big\{(i,j,k): M_\Omega(i,j,k)=1\big\}
\]
the corresponding admissible measurement set for the grid-wide QR formulation. In the direct QR experiments below, $\Omega=\Omega_{\mathrm{grid}}$ contains admissible spatial--property locations. The separate well-only study in Sec.~\ref{subsec:well_based_measurements} does not define a second QR admissible set over wells; instead, it restricts the measurement stage to subsets $\mathcal{W}_N\subseteq\mathcal{W}$ of physical wells. For the grid-wide QR factorization itself, we also let $c(i,j,k)\in\{1,\dots,IJK\}$ denote the column of $\mathcal{A}_{(4)}$ associated with index $(i,j,k)$.
\par
Formally, for the tensor dictionary \(\mathcal{A}\in\mathbb{R}^{I\times J\times K\times W}\), we apply QR with column pivoting to its mode-4 unfolding,
\begin{equation}
\mathcal{A}_{(4)}\,\Pi \;=\; Q R,
\label{eq:tbmd_projection_fixed}
\end{equation}
Equation~\eqref{eq:tbmd_projection_fixed} defines the QR factorization used for pivot selection, where \(\Pi\in\mathbb{R}^{(IJK)\times(IJK)}\) is the column-permutation matrix, \(Q\in\mathbb{R}^{W\times W}\) is orthogonal, and \(R\in\mathbb{R}^{W\times(IJK)}\) is upper-trapezoidal. The first \(N\) pivot columns of \(\Pi\) define the ordered sensor set
\[
\Omega_N = \{(i_1,j_1,k_1),\dots,(i_N,j_N,k_N)\}\subseteq \Omega.
\]
The QR-based sensor budget is constrained by both the dictionary depth and the number of admissible measurement locations:
\[
N \leq \min(W,|\Omega|),
\]
where $W$ denotes the number of retained dictionary fibers along the last tensor mode and $\Omega$ is the admissible grid-wide measurement set. In the reported QR sweeps, $\Omega=\Omega_{\mathrm{grid}}$ contains admissible spatial--property locations. The separate well-only experiments do not rerun QR on a well-restricted mask; instead, they restrict the TBCS measurement operator to subsets $\mathcal{W}_N\subseteq\mathcal{W}$, with $|\mathcal{W}|=30$ physical wells and budget $N=|\mathcal{W}_N|$.
At iteration $d$, admissible columns are scored by the \(\ell_1\)-norm of their \emph{trailing} residual,
\begin{equation}
\rho_d(i,j,k) \;=\; \|R_{d:W,\,c_d(i,j,k)}\|_1
\;=\; \|\mathcal{R}_{ijk,\,d:W}\|_1,
\label{eq:l1_norm_tube_fixed}
\end{equation}
where $c_d(i,j,k)$ denotes the current column position of $(i,j,k)$ after the first $d-1$ swaps and $\mathcal{R}=\operatorname{fold}_4(R)$. The pivot with maximal $\rho_d$ is swapped into position $d$, the permutation vector is updated explicitly, and a Householder reflector is applied to the trailing block $R_{d:W,\,d:IJK}$. If the maximal score vanishes, the remaining admissible columns add no further independent information and the procedure terminates.
\par
In the original article by Zhong et al.~\cite{zhong2024tbmd}, the same idea is presented for a third-order setting. Here we preserve the reflector logic while writing the pivoting explicitly as a column permutation of the unfolded fourth-order dictionary. Algorithm~\ref{alg:mode4_qr_algo} is stated for a generic availability mask. In the numerical results below, however, direct QR-selected sets $\Omega_N$ and the associated budget sweeps are reported only for the grid-wide admissible set $\Omega_{\mathrm{grid}}$; Sec.~\ref{subsec:well_based_measurements} instead uses random subsets $\mathcal{W}_N\subseteq\mathcal{W}$ and does not report a separate QR reranking within a well-only mask.

\begin{algorithm}[htbp]
\caption{Mode-4 fiber–pivot QR on $\mathcal{A}_{(4)}$ with availability mask and $\ell_1$ criterion}
\label{alg:mode4_qr_algo}
\begin{algorithmic}
\State \textbf{Input:} $\mathcal{A}\in\mathbb{R}^{I\times J\times K\times W}$, target number of sensors $N\le \min(W,|\Omega|)$; availability mask $M_\Omega$ with $\Omega=\{(i,j,k): M_\Omega(i,j,k)=1\}$.

\State \textbf{Initialization:} $R \leftarrow \mathcal{A}_{(4)}$, $Q \leftarrow I_W$, $\Omega_{\mathrm{adm}} \leftarrow \Omega$, $\Omega_N \leftarrow \varnothing$.
\State $\pi \leftarrow [1,\dots,IJK]$ and $\nu_j \leftarrow c^{-1}(j)$ for $j=1,\dots,IJK$.
\For{$d=1,\dots,N$}
  \State choose $j^{*}=\arg\max\limits_{\substack{j\ge d\\ \nu_j\in\Omega_{\mathrm{adm}}}}\|R_{d:W,j}\|_1$;
  \State $\rho^{*}\leftarrow\|R_{d:W,j^{*}}\|_1$;
  \If{$\rho^{*}=0$}
    \State \textbf{break};
  \EndIf
  \State swap columns $d$ and $j^{*}$ of $R$; swap $\pi_d\leftrightarrow\pi_{j^{*}}$ and $\nu_d\leftrightarrow\nu_{j^{*}}$;
  \State append $\nu_d$ to $\Omega_N$ and remove $\nu_d$ from $\Omega_{\mathrm{adm}}$;
  \State $t\leftarrow R_{d:W,d}$;\quad $\sigma\leftarrow\|t\|_2$;
  \If{$\sigma=0$}
    \State \textbf{break};
  \EndIf
  \State $s\leftarrow 1$ if $t_1\ge 0$, else $s\leftarrow -1$;
  \State $v\leftarrow t+s\,\sigma\,e_1$;\quad $u\leftarrow v/\|v\|_2$;
  \State $R_{d:W,d:IJK}\leftarrow R_{d:W,d:IJK}-2u\bigl(u^{\top}R_{d:W,d:IJK}\bigr)$;
  \State $Q_{:,d:W}\leftarrow Q_{:,d:W}-2\bigl(Q_{:,d:W}u\bigr)u^{\top}$;
\EndFor
\State form $\Pi$ from the final permutation vector $\pi$.
\State \textbf{Output:} $\Omega_N,\pi,\Pi,Q,R$.
\end{algorithmic}
\end{algorithm}

\noindent
\emph{Remarks.} 
(i) The availability mask $M_\Omega$ defines the fixed feasible set $\Omega$, whereas $\Omega_{\mathrm{adm}}$ is the mutable working copy used inside Algorithm~\ref{alg:mode4_qr_algo}.   
(ii) The permutation vector $\pi$ records the actual column swaps; the associated matrix \(\Pi\) satisfies $\mathcal{A}_{(4)}\Pi=QR$.  
(iii) The stopping rule $\rho^{*}=0$ (equivalently, $\sigma=0$) handles rank-deficient or exhausted admissible sets and avoids an undefined Householder vector.  
(iv) The measurement stage in Sec.~\ref{subsec:tensor_cs} uses $\Omega_N$ (or its binary mask $M_{\Omega_N}$), not the permutation matrix \(\Pi\); \(\Pi\) is retained only to state the QR factorization itself.

\subsection{Tensor Compressive Sensing and Field Reconstruction}
\label{subsec:tensor_cs}
\par
Following sensor selection, we reconstruct held-out 3D reservoir snapshots from sparse observations by formulating a compressive sensing problem~\cite{donoho2006,candes2007} over the tensor dictionary. This framework extends $\ell_1$-norm minimization to multidimensional data~\cite{caiafa2013multidimcs} and naturally accommodates joint (block-sparse) recovery across multiple properties~\cite{eldar2010_block}. In the present manuscript, the joint pressure+saturation recovery is carried out on the normalized tensor $\widetilde{\mathcal{X}}$ defined in Sec.~\ref{subsec:conventions_units}; for notational simplicity, the tildes are omitted in this subsection. In the reported experiments, TBCS is applied independently to each held-out time step. For a held-out index $\ell$, let $\mathcal{Z}^{(\ell)}\in\mathbb{R}^{I\times J\times K}$ denote the 3D snapshot defined by $\mathcal{Z}^{(\ell)}_{ijk}=\mathcal{X}_{ijk\ell}$. A persistent grid-wide sensor location $q=(i_q,j_q,k_q)\in\Omega_{\mathrm{grid}}$ then yields the scalar snapshot measurement
\[
y_q^{(\ell)}=\mathcal{Z}^{(\ell)}_{i_q j_q k_q}=\mathcal{X}_{i_q j_q k_q \ell}.
\]
The temporal fiber $\mathcal{X}_{i_q j_q k_q:}$ is therefore the sequence $\{y_q^{(\ell)}\}_{\ell=1}^{L}$ of such scalar measurements, not a separate object passed to the TBCS solver. For readability, the superscript $(\ell)$ is omitted below whenever the recovery problem is written for one snapshot. Consequently, $\mathcal{A}$, $A_N$, and $\mathbf{y}_N$ below denote the training-fitted property-wise min--max normalized dictionary, sensing matrix, and measurements for a single snapshot. This fixes a common dimensionless numerical scale for the inverse problem; when physical-unit fields are needed for interpretation, the inverse property-wise affine map is applied after recovery. Let $\mathcal{A}\in\mathbb{R}^{I\times J\times K\times W}$ be the stack of $W$ TBMD modes and $\mathbf{x}\in\mathbb{R}^{W}$ the modal coefficients. In the grid-wide experiments, the active set is $\Omega_N=\{(i_r,j_r,k_r)\}_{r=1}^N\subseteq\Omega_{\mathrm{grid}}$, and we define the measurement operator $\Phi_{\Omega_N}:\mathbb{R}^{I\times J\times K}\to\mathbb{R}^{N}$ by
\[
\Phi_{\Omega_N}(\mathcal{Z})
= \big[\mathcal{Z}_{i_1j_1k_1},\dots,\mathcal{Z}_{i_Nj_Nk_N}\big]^{\top},
\]
which returns one scalar per selected spatial--property channel. For the well-only setting, let $\mathcal{W}_N=\{w_1,\dots,w_N\}\subseteq\mathcal{W}$ denote the selected physical wells, so the budget remains $N=|\mathcal{W}_N|$. If both properties are used, a selected well $w$ at parent cell $(i_w,j_w)$ provides the property vector
\[
\mathbf{z}_w
=
\big[\mathcal{Z}_{i_w j_w 1},\ldots,\mathcal{Z}_{i_w j_w K}\big]^{\top}.
\]
Under the fixed convention of Sec.~\ref{subsec:conventions_units}, the present two-property case stacks pressure first and oil saturation second.
Thus, the stacked well-only measurement operator is
\[
\Psi_{\mathcal{W}_N}(\mathcal{Z})
=
\big[\mathbf{z}_{w_1}^{\top},\ldots,\mathbf{z}_{w_N}^{\top}\big]^{\top},
\qquad
\Psi_{\mathcal{W}_N}:\mathbb{R}^{I\times J\times K}\to\mathbb{R}^{N K},
\]
so in the joint pressure+saturation case considered here ($K=2$), each selected physical well contributes two stacked measurements while the admissible well set size remains $|\mathcal{W}|=30$ physical wells rather than $30\times2$ property-specific channels. In the single-property well-only case, the same construction reduces to one scalar measurement per selected well, i.e., effectively $K=1$ and $m=N$. Let $\mathcal{M}_N$ denote the relevant measurement operator, i.e., $\mathcal{M}_N=\Phi_{\Omega_N}$ in the grid-wide case and $\mathcal{M}_N=\Psi_{\mathcal{W}_N}$ in the well-only multi-property case, and let $\mathbf{y}_N\in\mathbb{R}^{m}$ be the corresponding measurement vector, where $m=N$ for the grid-wide formulation and $m=NK$ for the well-only multi-property formulation. The strict formulation reads
\begin{equation}
\min_{\mathbf{x}} \;\|\mathbf{x}\|_1
\quad \text{s.t.} \quad
\mathbf{y}_{N} \;=\; \mathcal{M}_{N}\!\big(\mathcal{A} \times_{(4)} \mathbf{x}\big),
\label{eq:l1_min_problem}
\end{equation}
and its relaxed version
\begin{equation}
\min_{\mathbf{x}} \;\varepsilon \|\mathbf{x}\|_1
\;+\; \frac{1}{2}\, \big\| \mathcal{M}_{N}\!\big(\mathcal{A} \times_{(4)} \mathbf{x}\big) - \mathbf{y}_{N} \big\|_2^2,
\label{eq:l1_l2_problem}
\end{equation}
with $\varepsilon>0$ balancing sparsity and data fit. Equations~\eqref{eq:l1_min_problem} and~\eqref{eq:l1_l2_problem} define the exact and relaxed sparse recovery problems, respectively. This is solved via ADMM by introducing an auxiliary variable $\mathbf{d}$ and scaled dual variable $\mathbf{p}$.
Once $\hat{\mathbf{x}}$ is obtained for a given snapshot, the reconstructed field is
\[
\hat{\mathcal{Z}}=\mathcal{A}\times_{(4)}\hat{\mathbf{x}}.
\]
Repeating this recovery over the held-out time indices yields the reconstructed temporal trajectory.
\par
Let $A_{N}\in\mathbb{R}^{m\times W}$ be the sensing matrix induced by $\mathcal{M}_{N}$. In the grid-wide case, its $r$-th row is $\mathcal{A}_{i_rj_rk_r:}^{\top}$. In the well-only multi-property case, $A_N$ is obtained by stacking the rows $\mathcal{A}_{i_{w_r}j_{w_r}k:}^{\top}$ for $r=1,\dots,N$ and $k=1,\dots,K$, in the same order as in $\Psi_{\mathcal{W}_N}(\mathcal{Z})$. In the present two-property case, each well therefore contributes the pressure row first and the oil-saturation row second, i.e.,
\[
A_N
=
\begin{bmatrix}
\mathcal{A}_{i_{w_1}j_{w_1}1:}^{\top}\\
\mathcal{A}_{i_{w_1}j_{w_1}2:}^{\top}\\
\vdots\\
\mathcal{A}_{i_{w_N}j_{w_N}1:}^{\top}\\
\mathcal{A}_{i_{w_N}j_{w_N}2:}^{\top}
\end{bmatrix}.
\]
Define the soft-thresholding operator
\[
\operatorname{soft}(z,\kappa) \;=\; \operatorname{sign}(z)\,\max(|z|-\kappa,0),
\]
and the relaxed iterate
\[
\hat{\mathbf{x}}^{(n+1)} \;=\; \lambda_{\text{relax}}\,\mathbf{x}^{(n+1)} + (1-\lambda_{\text{relax}})\,\mathbf{d}^{(n)},
\qquad \lambda_{\text{relax}}\in(0,1).
\]
The ADMM scheme is summarized in Algorithm~\ref{alg:tbmd_cs_admm}.

\begin{algorithm}[htbp]
\caption{ADMM for an $\ell_1$-regularized problem (see also \cite{boyd2011_admm})}
\label{alg:tbmd_cs_admm}
\small
\begin{algorithmic}
\State \textbf{Input:} sensing matrix $A_N\in\mathbb{R}^{m\times W}$; measurements $\mathbf{y}_{N}\in\mathbb{R}^{m}$;
\State sparsity weight $\varepsilon>0$, initial penalty $\delta^{(0)}>0$, maximum penalty $\delta_{\max}$,
\State residual imbalance threshold $\mu>1$, update factor $\eta>1$, tolerance \texttt{tol}.
\State \textbf{Initialization:} $\mathbf{x}^{(0)} \leftarrow 0$,\ $\mathbf{d}^{(0)} \leftarrow 0$,\ $\mathbf{p}^{(0)} \leftarrow 0$.
\For{$n = 0,1,2,\dots$ \textbf{until} the stopping criterion is met}
  \State \textbf{(x-step)} Solve the normal equations:
  \State $\big(A_{N}^{\top} A_{N} + \delta^{(n)} I\big)\,\mathbf{x}^{(n+1)} = A_{N}^{\top} \mathbf{y}_{N} + \delta^{(n)}\big(\mathbf{d}^{(n)}-\mathbf{p}^{(n)}\big)$
  \State \textbf{(relaxation)} $\hat{\mathbf{x}}^{(n+1)} \leftarrow \lambda_{\text{relax}}\,\mathbf{x}^{(n+1)} + (1-\lambda_{\text{relax}})\,\mathbf{d}^{(n)}$, \ \ $\lambda_{\text{relax}}\!\in\!(0,1)$
  \State \textbf{(d-step)} Soft-thresholding with $\kappa^{(n)}=\varepsilon/\delta^{(n)}$:
  \State $\mathbf{d}^{(n+1)} \leftarrow \operatorname{soft}\!\big(\hat{\mathbf{x}}^{(n+1)}+\mathbf{p}^{(n)},\ \kappa^{(n)}\big)$
  \State \textbf{(dual update)} $\mathbf{p}^{(n+1)} \leftarrow \mathbf{p}^{(n)} + \big(\hat{\mathbf{x}}^{(n+1)}-\mathbf{d}^{(n+1)}\big)$
  \State \textbf{Compute residuals:} $r^{(n+1)}=\hat{\mathbf{x}}^{(n+1)}-\mathbf{d}^{(n+1)}$, \ \ $s^{(n+1)}=\delta^{(n)}\big(\mathbf{d}^{(n+1)}-\mathbf{d}^{(n)}\big)$
  \If{$\max\!\big(\|r^{(n+1)}\|_2,\ \|s^{(n+1)}\|_2\big) < \texttt{tol}$}
    \State \textbf{break}
  \EndIf
  \If{$\|r^{(n+1)}\|_2 > \mu\|s^{(n+1)}\|_2$}
    \State $\delta^{(n+1)} \leftarrow \min\big(\eta\,\delta^{(n)},\,\delta_{\max}\big)$
  \ElsIf{$\|s^{(n+1)}\|_2 > \mu\|r^{(n+1)}\|_2$}
    \State $\delta^{(n+1)} \leftarrow \delta^{(n)}/\eta$
  \Else
    \State $\delta^{(n+1)} \leftarrow \delta^{(n)}$
  \EndIf
  \State \textbf{(dual rescaling)} $\mathbf{p}^{(n+1)} \leftarrow \dfrac{\delta^{(n)}}{\delta^{(n+1)}}\,\mathbf{p}^{(n+1)}$
\EndFor
\State \textbf{Output:} estimate of modal coefficients $\hat{\mathbf{x}}=\mathbf{x}^{(n+1)}$.
\end{algorithmic}
\end{algorithm}

\noindent\emph{Notes.}
The x-step (normal equations) enforces data fidelity and the augmented Lagrangian penalty; the d-step promotes sparsity via $\ell_1$-shrinkage; the dual update $\mathbf{p}$ drives consensus between $\mathbf{x}$ and $\mathbf{d}$. Primal and dual residuals $r^{(n)}$ and $s^{(n)}$ monitor convergence. The objective function is
\[
J \;=\; \tfrac{1}{2}\|A_{N}\mathbf{x}-\mathbf{y}_{N}\|_2^2 + \varepsilon\|\mathbf{d}\|_1.
\]
The penalty update follows the residual-balancing rule specified in the released configuration, with a maximum-penalty cap. Because $\mathbf{p}$ is the scaled dual variable, it is rescaled as $\mathbf{p}^{(n+1)}\leftarrow (\delta^{(n)}/\delta^{(n+1)})\,\mathbf{p}^{(n+1)}$ after each penalty change so that the corresponding unscaled multiplier remains unchanged; setting the update factor to one recovers the fixed-penalty ADMM variant.

\section{Proposed Method}
\label{subsec:integrated_approach}
\par
\begin{figure}[htbp]
\centering
\resizebox{\textwidth}{!}{%
\begin{tikzpicture}[
  font=\small, >=Latex, node distance=5mm and 8mm,
  block/.style={draw,rounded corners,fill=gray!8,inner sep=4pt,align=center},
  tensor/.style={draw,thick,minimum width=22mm,minimum height=14mm,fill=blue!5,align=center},
  dict/.style={draw,thick,minimum width=22mm,minimum height=11mm,fill=orange!10,align=center},
  meas/.style={draw,minimum width=22mm,minimum height=10mm,fill=teal!8,align=center},
  arrow/.style={-Latex,thick}
]
\node[tensor,label=below:{\scriptsize (a) $\mathcal{X}$}] (X)
  {$\mathcal{X}\in\mathbb{R}^{I\times J\times K\times L}$\\
   {\scriptsize $K=2$, fixed property order}};
\node[block,right=of X,label=below:{\scriptsize (b) TBMD}] (hosvd)
  {$\mathcal{X}\approx \mathcal{G}\times_1 U^{(1)}\times_2 U^{(2)}\times_3 U^{(3)}\times_4 U^{(4)}$};
\node[dict,right=14mm of hosvd,label=below:{\scriptsize (c) Dict. $\mathcal{A}$}] (A)
  {$\mathcal{A}=\mathcal{G}\times_1 U^{(1)}\times_2 U^{(2)}\times_3 U^{(3)}$\\
   {\scriptsize $W$ modes}};
\node[block,right=of A,text width=48mm,align=center,label=below:{\scriptsize (d) Mode-4 fiber QR}] (qr)
  {$\mathcal{A}_{(4)}\Pi = Q R$\\[2pt]
   {\scriptsize first $N$ pivot columns $\Rightarrow \Omega_N\subseteq\Omega$}};
\node[meas,below=10mm of A,label=below:{\scriptsize (e) Measurements}] (Y)
  {$\mathbf{y}_{N}\in\mathbb{R}^{m}$\\ {\scriptsize sparse snapshot measurements}};
\node[block,fill=red!6,label=below:{\scriptsize sensors}] at (Y -| qr) (sens)
  {{\footnotesize selected set $\Omega_N$}\\{\footnotesize permutation $\Pi$}};
\node[block,left=of Y,text width=54mm,align=center,label=below:{\scriptsize (f) T-CS (ADMM)}] (cs)
  {$\min_{\mathbf{x}}\ \varepsilon\|\mathbf{x}\|_{1} + \tfrac12\|\mathcal{M}_{N}(\mathcal{A} \times_{(4)} \mathbf{x}) - \mathbf{y}_{N}\|_2^2$};
\node[tensor,left=of cs,label=below:{\scriptsize (g) Reconstruction}] (Xhat)
  {$\hat{\mathcal{Z}}=\mathcal{A}\times_{(4)} \hat{\mathbf{x}}$};
\draw[arrow] (X) -- (hosvd);
\draw[arrow] (hosvd) -- (A);
\draw[arrow] (A) -- (qr);
\draw[arrow] (qr) -- (sens);
\draw[arrow] (sens) -- (Y);
\draw[arrow] (A.south) -- (Y.north);
\draw[arrow] (Y) -- (cs);
\draw[arrow] (cs) -- (Xhat);
\end{tikzpicture}%
}
\caption{Integrated TBMD--QR--TBCS workflow and the two measurement operators used for grid-wide and well-only reconstruction.}
\label{fig:framework_schematic}
\end{figure}
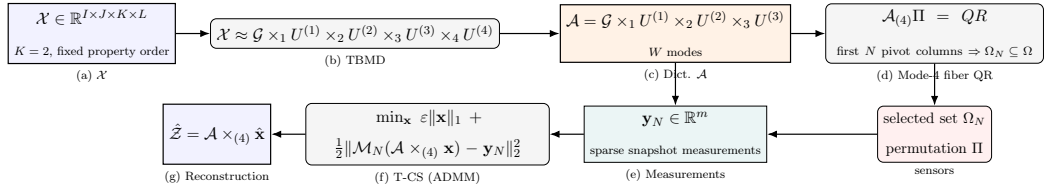

The integrated pipeline $\mathrm{TBMD}\to\mathrm{QR}\to\mathrm{TBCS}$ provides a structured methodological workflow for sparse monitoring and reconstruction under the tested Brugge protocol:
\begin{enumerate}
  \item \textbf{TBMD (Sec.~\ref{subsec:tbmd})} extracts a compact modal dictionary $\mathcal{A}\in\mathbb{R}^{I\times J\times K\times W}$ with dictionary depth $W$ and controlled energy loss from the property-wise normalized training tensor, retaining gridded spatial organization and cross-property structure via HOSVD.
  \item \textbf{Mode-4 fiber-pivot QR (Sec.~\ref{subsec:tensor_qr})} ranks spatially distributed fibers $(i,j,k)$ according to the trailing $\ell_1$-residual criterion~\eqref{eq:l1_norm_tube_fixed}, honoring the availability mask $M_\Omega$. It yields a permutation vector $\pi$ and associated column permutation $\Pi$ on $\mathcal{A}_{(4)}$; the first $N$ pivots define the selected sensor set $\Omega_N$ subject to $N\leq\min(W,|\Omega|)$. In the reported results, direct QR is evaluated on $\Omega_{\mathrm{grid}}$, whereas the well-only study uses subsets $\mathcal{W}_N\subseteq\mathcal{W}$ only as a budget-constrained measurement restriction. See Algorithm~\ref{alg:mode4_qr_algo}.
  \item \textbf{TBCS via ADMM (Sec.~\ref{subsec:tensor_cs})} recovers snapshot-wise modal coefficients $\mathbf{x}\in\mathbb{R}^W$ from sparse normalized measurements $\mathbf{y}_{N}$ via the appropriate measurement operator ($\Phi_{\Omega_N}$ for grid-wide sensing or $\Psi_{\mathcal{W}_N}$ for stacked well-only multi-property sensing), alternating x-step, d-step, and dual update; the normalized snapshot $\hat{\mathcal{Z}}=\mathcal{A}\times_{(4)}\hat{\mathbf{x}}$ is then reconstructed and, if needed, mapped back to physical units property-wise. See Algorithm~\ref{alg:tbmd_cs_admm}.
\end{enumerate}
Practically, this yields: (i)~systematic dimension reduction while retaining gridded spatial organization and cross-property structure; (ii)~QR-guided layouts that highlight response-contrast regions in the tested data; (iii)~internally consistent reconstruction of pressure and saturation fields from limited, spatially distributed measurements under the tested settings; and (iv)~a workflow structure that can be evaluated in future time-constrained monitoring studies after dedicated runtime validation. A schematic overview of this integrated workflow and the sensing regimes is provided in Figure~\ref{fig:framework_schematic}.
In the reported experiments, direct QR-selected layouts are studied only on the grid-wide admissible set. The existing-well study in Sec.~\ref{subsec:well_based_measurements} instead evaluates the same reconstruction stage under well-constrained measurement subsets, and Sec.~\ref{subsec:well_significance} maps the grid-wide QR ranking onto the well inventory for operational interpretation.
\par
Together, the TBMD--QR--TBCS pipeline forms an interpretable computational workflow for reservoir monitoring studies, while enabling sparse-measurement field reconstruction under the tested settings without making a runtime or deployment claim.

\FloatBarrier
\FloatBarrier

\section{Algorithmic Implementation}
\label{sec:algorithmic_implementation}
The implementation is organized run-wise. For each of the 10 control realizations, the first 80\% of snapshots are used to fit the property-wise min--max normalization parameters, compute the truncated TBMD basis, build the tensor dictionary, and construct the grid-wide QR ranking. The remaining 20\% of snapshots are reconstructed independently by TBCS and are used only for held-out evaluation. The run index is therefore an outer loop rather than an additional tensor mode.
\par
Algorithm~\ref{alg:mode4_qr_algo} takes the dictionary $\mathcal{A}\in\mathbb{R}^{I\times J\times K\times W}$ and a feasible mask $M_\Omega$ as inputs and returns an ordered grid-wide sensor set $\Omega_N$. Algorithm~\ref{alg:tbmd_cs_admm} takes the corresponding sensing matrix $A_N$, the measurement vector $\mathbf{y}_N$, and ADMM parameters as inputs and returns the modal coefficients used to synthesize $\hat{\mathcal{Z}}$. In the grid-wide experiments, $A_N$ is generated from QR-selected spatial--property locations. In the well-only experiments, $A_N$ is generated from selected physical wells and the relevant property channels; no separate QR reranking over the well set is reported.
\par
The reported grid-wide budget sweeps use dictionary depth $W=480$, whereas selected qualitative map reconstructions use a 200-mode dictionary only for visualization. The well-only tests use random subsets of the 30 existing wells, with the all-well case corresponding to $N=30$. Repository-derived reproducibility parameters are summarized in Sec.~\ref{subsec:reproducibility_summary}. Detailed hardware-specific scalability claims and wall-clock runtime measurements are omitted to maintain focus on the mathematical and algorithmic formulation.

\subsection{Reproducibility Summary}
\label{subsec:reproducibility_summary}

The implementation was written in Python using the \texttt{tbmd} package version 2.0.0. The Brugge tensors were loaded from \texttt{data\_exp\_4\_.h5}, and well coordinates were loaded from \texttt{all\_wells\_exp\_4.json}. The data were split sequentially into training and testing subsets using an 80/20 split without random shuffling. Property-wise min--max normalization was estimated from the training data using separate parameters for each property and applied to both training and test tensors. No explicit background value was used.
\par
The fixed random seed was set to 0 for TensorLy, NumPy, TBMD decomposition, and QR-based sensor selection. The TBMD/Tucker decomposition used multilinear ranks $[48,48,2,48]$ and tolerance $\epsilon=10^{-2}$ with SVD-based Tucker initialization. QR-based sensor placement used $N=200$ candidate sensor locations, random state 0, orthogonality checking enabled, and no uniform-distribution constraint.
\par
The TBCS recovery was solved by ADMM with maximum iteration count 1000, stopping tolerance $10^{-4}$, L1 regularization weight $\epsilon_{\ell_1}=10^{-2}$, initial penalty $\delta_0=1.0$, maximum penalty $\delta_{\max}=1.0$, and relaxation parameter $\lambda=0.95$. The ADMM x-update used a Cholesky-based linear solver with diagonal regularization $10^{-8}$. The implementation uses the penalty-update rule specified in the released configuration: Boyd-style residual balancing with imbalance threshold 10, growth factor 2, shrink factor 1/2, and maximum-penalty cap. The TBCS stage was configured to run on CPU using \texttt{torch.float32}. The TBMD and modal-processing stages were configured to use the PyTorch MPS backend where available.
\par
The software environment used Python $\geq$3.10 with NumPy 1.26.4, SciPy 1.11.4, PyTorch 2.3.1, TensorLy 0.9.0, scikit-learn 1.5.2, pandas 2.2.2, h5py $\geq$3.12.1, scikit-image 0.23.2, and Matplotlib 3.8.4. Table~\ref{tab:runtime_hardware} summarizes the available computational environment details.

\begin{table}[t]
\centering
\caption{Computational environment and hardware summary for the reported experiments.}
\label{tab:runtime_hardware}
\begin{tabular}{ll}
\toprule
Item & Value \\
\midrule
Processor & Apple M2 Pro (12-core) \\
Memory & 16 GB Unified RAM \\
OS & macOS \\
Python version & $\geq$3.10 \\
Main package version & tbmd 2.0.0 \\
TBMD / Modal processing backend & PyTorch MPS \\
TBCS backend & CPU (\texttt{torch.float32}) \\
Peak memory (TBMD) & $\approx$ 31 MB \\
Peak memory (QR + CS per snapshot) & $\approx$ 2 MB \\
Runtime (TBMD decomposition) & $\approx$ 0.2 s \\
Runtime (QR + CS per snapshot) & $\approx$ 0.01 s \\
\bottomrule
\end{tabular}
\end{table}
\section{Computational Complexity}
\label{sec:computational_complexity}
Let $M=IJK$ denote the number of candidate spatial--property locations, $W$ the retained dictionary depth, $N$ the sensor budget, $m$ the number of scalar measurements in one snapshot-level inverse problem, and $n_{\mathrm{ADMM}}$ the number of ADMM iterations. A dense HOSVD fit is dominated by truncated singular-value decompositions of the tensor unfoldings. For an exact dense SVD of the mode-$n$ unfolding with size $I_n\times J_n$, the worst-case cost scales as $O(\min\{I_n^2J_n,I_nJ_n^2\})$; practical truncated or randomized SVD implementations can reduce this cost when the target multilinear ranks are small.
\par
The mode-4 QR stage applies pivoting to $\mathcal{A}_{(4)}\in\mathbb{R}^{W\times M}$. A full dense pivoted QR factorization scales as $O(WM\min\{W,M\})$ with memory $O(WM)$. The budgeted implementation in Algorithm~\ref{alg:mode4_qr_algo} can terminate after $N$ pivots, giving a leading-order cost proportional to $O(NWM)$ for repeated residual scoring and Householder updates, up to implementation-dependent constants.
\par
For one held-out snapshot, TBCS solves an $m\times W$ sparse-regression problem. If the normal-equation factorization in the ADMM x-step is recomputed, the setup cost is $O(mW^2+W^3)$; if the active sensor set is fixed over many snapshots, this factorization can be reused and the per-iteration cost is dominated by matrix--vector products and triangular solves, approximately $O(mW+W^2)$. The total reconstruction cost scales linearly with the number of held-out snapshots and with the number of independently processed control realizations. The present complexity discussion is asymptotic, focusing on computational scaling rather than a hardware-specific latency assessment.

\section{Numerical Experiments}
\label{sec:dataset_description}
The experiments below are designed to evaluate transferability and internal consistency of the reservoir-specific extension, not to repeat the complete baseline suite from the previous TBMD study by Zhong et al.~\cite{zhong2024tbmd}. That study reported comparative evidence against POD and DMD for modal decomposition and against POD-based sparse-sensor reconstruction algorithms on cylinder-wake, airfoil-vortex, sea-surface-temperature, eigenface, and turbulent-channel-flow datasets using reconstruction-accuracy and mode/energy-loss criteria. Here, we instead focus on the coupled pressure--saturation tensor formulation, grid-wide spatial--property QR ranking, existing-well measurement operators, and QR-to-well diagnostic mapping on the Brugge benchmark. Consequently, the results should not be interpreted as a claim that the present Brugge implementation performs better than POD/SVD, DMD, DEIM/Q-DEIM, random sensing, or learning-based alternatives under the same reservoir-monitoring conditions.

\subsection{Dataset and Experimental Protocol}
\label{subsec:dataset_protocol}
\par
The effectiveness of the integrated methodology was evaluated using the Brugge field reservoir simulation benchmark. The Brugge benchmark, developed by the Netherlands Organization for Applied Scientific Research (TNO), serves as a widely recognized test case in the literature for validating algorithms related to model order reduction, history matching, and production optimization~\cite{peters2010Brugge, TNO2009}.
\par
The Brugge model represents a semi-synthetic reservoir structure elongated from east to west, featuring a major fault along the northern boundary and an internal fault crossing the central dome. The total areal extent of the model is approximately \SI{10}{\kilo\metre\squared}. The stratigraphic sequence comprises four productive layers, each characterized by distinct depositional and petrophysical properties:
\begin{itemize}
    \item Schelde – fluvial channel facies characterized by high-permeability sand bodies interbedded with impermeable clay layers, forming an anisotropic flow network;
    \item Maas – shallow-marine facies with alternating sandstone beds and carbonate concretions, exhibiting moderate heterogeneity and partially reduced permeability;
    \item Waal – the primary reservoir interval, composed of laterally continuous sandy facies with excellent filtration–capacity characteristics, serving as the main production target;
    \item Schie – a thin, low-permeability layer with limited contribution to the overall production performance.
\end{itemize}
\par
\begin{figure}[htbp]
  \centering
  \includegraphics[width=\textwidth]{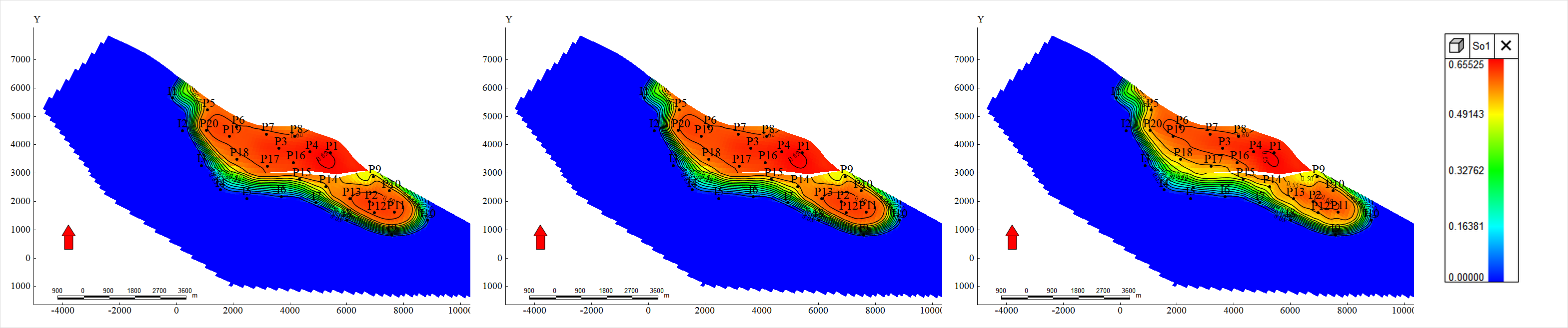}

  \includegraphics[width=\textwidth]{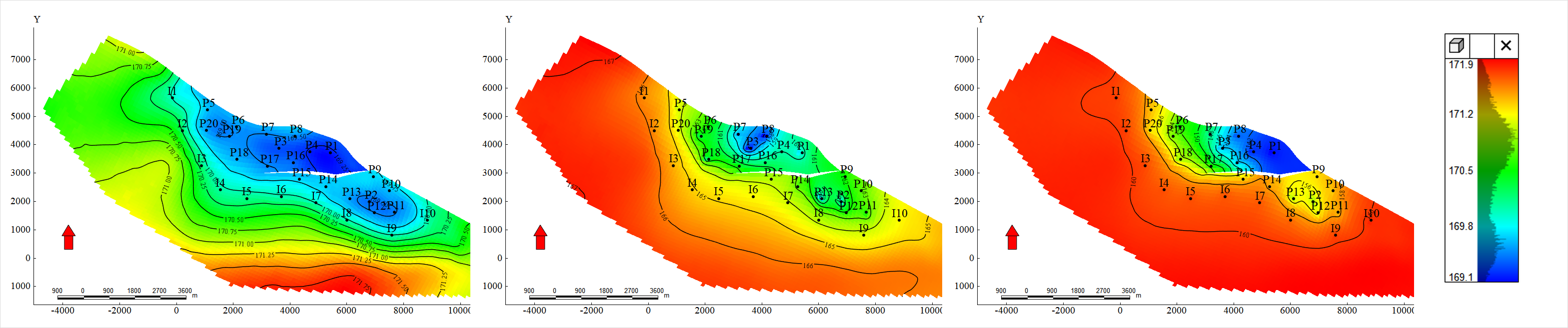}

  \caption{Brugge model fields arranged in two rows: oil saturation (top) and reservoir pressure (bottom) at three time steps $t_1$, $t_2$, $t_3$. For each property, a single enlarged color scale is used consistently across all three snapshots to improve readability and direct visual comparability. This row order is purely for visualization; the tensor formulation still follows the fixed property order defined in Sec.~\ref{subsec:conventions_units}.}
  \label{fig:brugge_fields}
\end{figure}

A major structural fault traverses the northern and central parts of the Brugge model, separating distinct flow compartments within the reservoir. 
This fault is visualized as a red line in Fig.~\ref{fig:brugge_fields} and in the subsequent map-based result figures, where it marks the main discontinuity in permeability and pressure fields.
\par
While the original TNO Brugge configuration assumes a standardized production schedule, the present study intentionally modifies the well-control scenario to evaluate the behavior of the proposed TBMD-based approach under non-standardized conditions. Specifically, 20 wells were operated as producers with randomly assigned bottom-hole pressures ranging from 60 to \SI{120}{bar} during the first 15 months of simulation. Subsequently, 10 wells were activated as injectors starting from the 15\textsuperscript{th} month, each maintaining a fixed injection pressure of \SI{175}{bar}, exceeding the initial reservoir pressure. This adjustment introduces more diverse boundary conditions, enabling an assessment of the method’s response to varying development strategies and dynamic conditions.
\par
The simulation covered a total period of 10 years (120 months), discretized into 134 time steps, providing sufficiently high temporal resolution for dynamic mode analysis. A set of 10 well-control scenarios was simulated, differing only in the randomly generated initial bottom-hole pressures of the producing wells. Importantly, the geological model remained identical across all runs, while only the boundary conditions varied. These 10 scenarios therefore constitute control realizations of operational settings under fixed geology, not geological realizations. Hence, the study evaluates control-induced dynamical variability rather than a geological ensemble. This setup allows a focused evaluation of dynamic variability driven by operational controls rather than geological uncertainty. 
\par
Each run $r\in\{1,\dots,10\}$ was structured as its own fourth-order tensor
\[
\mathcal{X}^{(r)}\in\mathbb{R}^{139 \times 48 \times 2 \times 134},
\]
where:
\begin{itemize}
    \item 139 – number of grid cells along the $x$-axis,
    \item 48 – number of grid cells along the $y$-axis,
    \item 2 – property mode with the fixed order defined in Sec.~\ref{subsec:conventions_units},
    \item 134 – number of time steps in the simulation.
\end{itemize}
Thus, the full study set is the collection $\{\mathcal{X}^{(r)}\}_{r=1}^{10}$, treated as 10 parallel study cases rather than as a single fifth-order tensor with an explicit run axis.
\par
Illustrative maps of the initial oil saturation and reservoir pressure fields are presented in Fig.~\ref{fig:brugge_fields}. Their top--bottom display order is purely visual and does not alter the fixed property order defined in Sec.~\ref{subsec:conventions_units}.
\par

\FloatBarrier
\FloatBarrier

Thus, each run captures the complete spatiotemporal evolution of pressure and saturation fields over the entire simulation period. The per-run data volume equals $139 \times 48 \times 2 \times 134 = 1{,}788{,}096 \approx 1.79 \times 10^{6}$ values; across all 10 well-control scenarios, the study set contains $17{,}880{,}960 \approx 1.79 \times 10^{7}$ values. Unless otherwise stated, all summary curves are obtained by evaluating each run on its own held-out 20\% time block and then aggregating the resulting run-wise metrics.
\par
The use of tensor-formatted data preserves the inherent spatial and temporal dependencies of the reservoir system, making the dataset particularly well suited for high-order tensor analysis. The primary objective of applying TBMD to this dataset is to extract dominant spatiotemporal modes of reservoir dynamics and to demonstrate accurate reconstruction of global fields from sparse measurements that realistically emulate field monitoring conditions.

\section{Numerical Results}
\label{sec:results}
\par
Unless stated otherwise, every metric curve in this section is obtained by applying the TBMD--QR--TBCS pipeline independently to each run tensor $\mathcal{X}^{(r)}$, evaluating on that run's held-out 20\% time steps, and then aggregating the resulting test metrics over $r=1,\dots,10$. The field maps shown below are representative held-out slices from individual runs rather than averages over runs.

\subsection{Field Reconstruction via TBMD: Full-Coverage Baseline and Grid-wide QR Sensing}
\label{subsec:full_coverage}
\subsubsection{Single-field TBMD Reconstruction (Pressure)}
\label{subsubsec:full_coverage_single}
\par
Initially, we evaluate the TBMD method's capacity for accurate field reconstruction under full observational coverage across all grid cells. This serves as a baseline verification of the model's capacity to approximate the original high-dimensional field using a reduced-order representation. The present subsection then continues with the direct grid-wide QR-selected sparse-sensing study on the same admissible grid. The distinct well-only existing-well experiment is reported separately in Sec.~\ref{subsec:well_based_measurements}.
\par
For this evaluation, pressure and oil saturation distributions from the Brugge reservoir benchmark model were considered over a sequence of time steps for each run. Each physical property (pressure and saturation) was initially analyzed separately, resulting in two distinct 3D tensors of dimensions $139 \times 48 \times 134$ (spatial grid $\times$ time) per run. Although the enhanced TBMD algorithm supports full 3D spatial input, for interpretability and visualization purposes, 2D spatial slices were analyzed for each time step.
\par
The temporal domain was split into training and test subsets within each run: the first 80\% (107 of 134 time steps) were used to compute TBMD modes (training), while the remaining 20\% (27 of 134 time steps) were reserved for reconstruction evaluation. TBMD decomposition was then applied to the training subset for each run-specific tensor to extract dominant spatiotemporal modes representing the principal dynamic patterns in the reservoir.
\par
The low-dimensional TBMD basis used for reconstruction was fitted on the training subset, ensuring a consistent modal representation and preventing any information leakage from the held-out test set.
All reconstruction metrics were then computed on the test subset.
\par
Before TBMD fitting, QR ranking, TBCS recovery, and metric evaluation, each physical property is normalized independently within each run. Specifically, for each property index $k\in\{1,2\}$ under the fixed convention of Sec.~\ref{subsec:conventions_units}, $A_{\min,\text{train}}^{(k)}$ and $A_{\max,\text{train}}^{(k)}$ are computed on the training subset (first 107 time steps), and the same affine transform is applied unchanged to both the training and test subsets:
\[
\widetilde{A} \;=\; \frac{A - A_{\min,\text{train}}^{(k)}}{A_{\max,\text{train}}^{(k)} - A_{\min,\text{train}}^{(k)}}\,.
\]
Hence the training portion of each property lies in $[0,1]$, while the held-out test portion is evaluated on the same fixed normalized scale with $\text{MAX}=1$. The TBMD basis, QR-selected dictionary, measurement vectors, and TBCS solver all operate on these normalized tensors; in the joint pressure+saturation setting, this prevents the data-fidelity term from being dominated by the pressure channel, so no additional residual weighting is introduced. SSIM is computed on the normalized maps using a Gaussian window $11\times 11$ with $\sigma=1.5$~\cite{wang2004ssim}.
\par
Reconstruction accuracy was estimated on the normalized held-out test set by comparing the original field $\widetilde{\mathbf{A}}_i$ with its reconstructed counterpart $\widetilde{\mathbf{A}}_i^{\text{re}}$ using several quantitative metrics.
\par
\noindent \textbf{Frobenius Error}
\[
\text{error}_i = \frac{\|\widetilde{\mathbf{A}}_i^{\text{re}} - \widetilde{\mathbf{A}}_i\|_F}{\|\widetilde{\mathbf{A}}_i\|_F}
\]
where $\|\cdot\|_F$ denotes the Frobenius norm.
\par
\noindent \textbf{Structural Similarity Index (SSIM)}
\[
\text{ssim}_i = \frac{(2\mu_{\widetilde{\mathbf{A}}_i^{\text{re}}} \mu_{\widetilde{\mathbf{A}}_i} + C_1)(2\sigma_{\widetilde{\mathbf{A}}_i^{\text{re}}\widetilde{\mathbf{A}}_i} + C_2)}{(\mu_{\widetilde{\mathbf{A}}_i^{\text{re}}}^2 + \mu_{\widetilde{\mathbf{A}}_i}^2 + C_1)(\sigma_{\widetilde{\mathbf{A}}_i^{\text{re}}}^2 + \sigma_{\widetilde{\mathbf{A}}_i}^2 + C_2)}
\]
where:
\begin{itemize}
    \item $\mu_{\widetilde{\mathbf{A}}_i}, \mu_{\widetilde{\mathbf{A}}_i^{\text{re}}}$ are mean intensities on the normalized scale,
    \item $\sigma_{\widetilde{\mathbf{A}}_i}^2, \sigma_{\widetilde{\mathbf{A}}_i^{\text{re}}}^2$ are the corresponding variances,
    \item $\sigma_{\widetilde{\mathbf{A}}_i^{\text{re}}\widetilde{\mathbf{A}}_i}$ is the covariance,
    \item $C_1, C_2$ are small regularization constants.
\end{itemize}
\par
\noindent \textbf{Peak Signal-to-Noise Ratio (PSNR)}
\[
\text{psnr}_i = 10 \cdot \log_{10} \left( \frac{\text{MAX}^2}{\text{MSE}_i} \right)
\]
\[
\text{MSE}_i = \frac{1}{n_{\mathrm{elem}}} \sum_{j=1}^{n_{\mathrm{elem}}} (\widetilde{\mathbf{A}}_i^{\text{re}}[j] - \widetilde{\mathbf{A}}_i[j])^2
\]
where $\text{MAX} = 1$ due to normalization and $n_{\mathrm{elem}}$ is the number of elements in the evaluated slice/tensor field.
\par
\begin{figure}[htbp]
    \centering
    
    \begin{subfigure}[t]{0.49\textwidth}
        \centering
        \includegraphics[width=\linewidth]{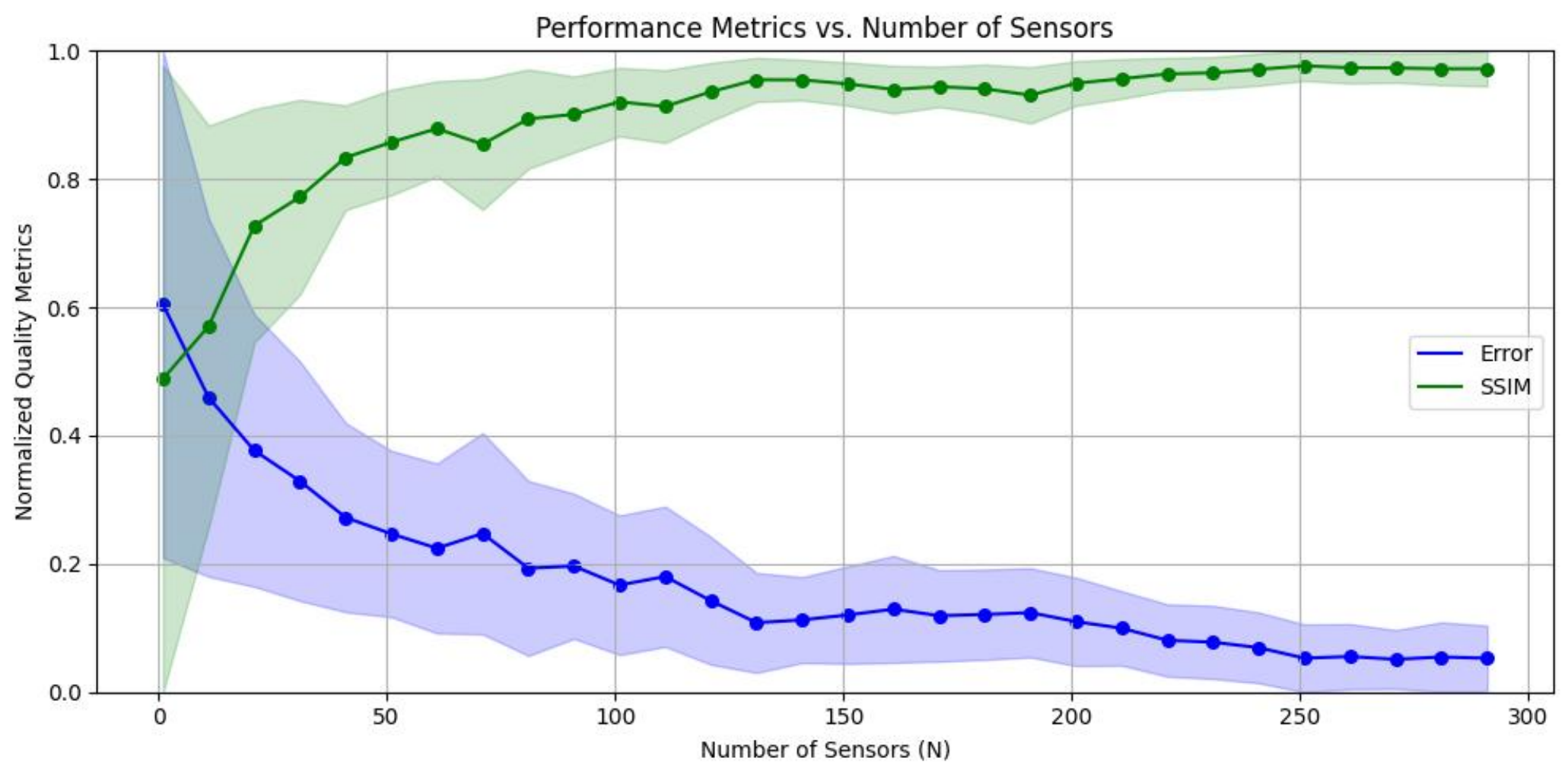}
        \caption{Relative Frobenius error and SSIM vs.\ number of QR-selected grid-wide sensor locations.}
        \label{fig:metric_convergence_a}
    \end{subfigure}\hfill
    \begin{subfigure}[t]{0.49\textwidth}
        \centering
        \includegraphics[width=\linewidth]{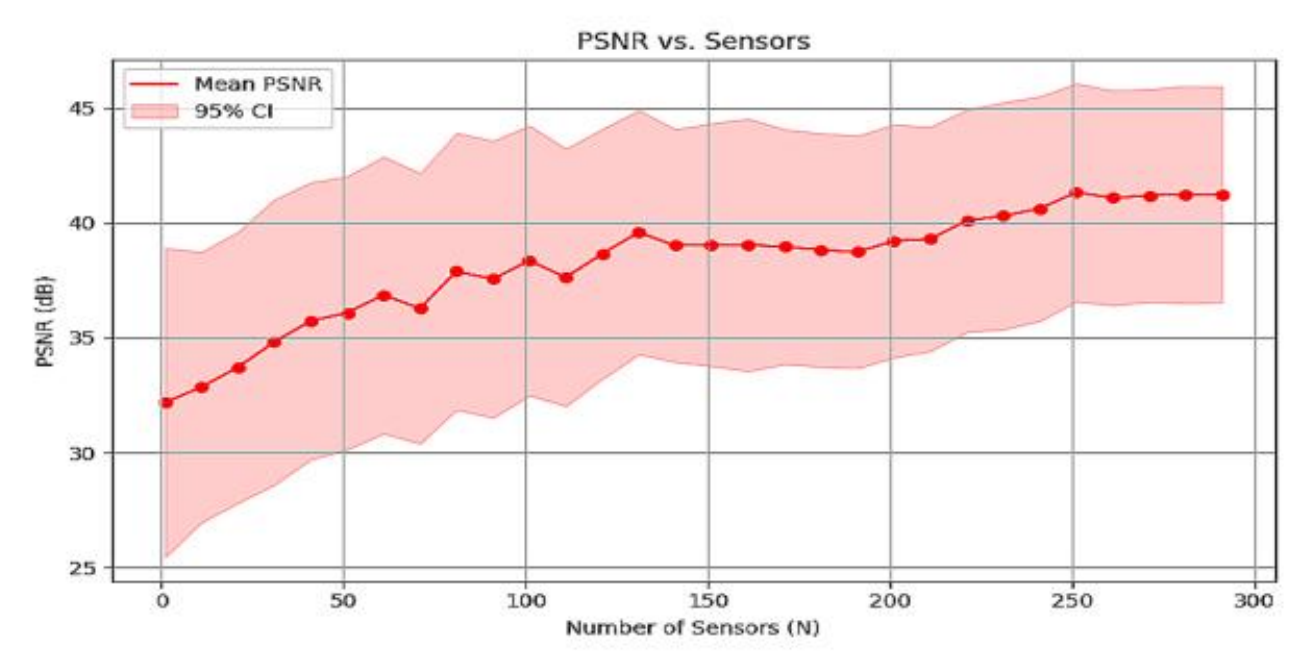}
        \caption{PSNR (dB) vs.\ number of QR-selected grid-wide sensor locations.}
        \label{fig:metric_convergence_b}
    \end{subfigure}
    
    \caption{Pressure reconstruction metrics for QR-selected grid-wide sensor budgets: (a) relative Frobenius error and SSIM; (b) PSNR.}
    \label{fig:metric_convergence}
\end{figure}

Figure~\ref{fig:metric_convergence}(a) illustrates how reconstruction quality improves as the sensor budget increases. In the grid-wide Brugge experiments, each sensor corresponds to a QR-selected grid-wide sensor location, i.e., a spatial--property grid cell at which pressure or saturation is measured. Throughout this grid-wide QR/TBCS study and the associated figures, $N$ denotes the number of QR-selected grid-wide sensor locations. The plotted mean curves aggregate held-out metrics from the sequential 80/20 split over the 10 control realizations with fixed geology. As $N$ increases, the error decreases rapidly, while the SSIM index rises and plateaus, indicating convergence of structural similarity. Figure~\ref{fig:metric_convergence}(b) presents the PSNR evolution, which likewise shows systematic improvement with additional QR-selected grid-wide sensor locations.
\par
\begin{figure}[htbp]
    \centering
    
    \begin{subfigure}[t]{0.49\textwidth}
        \centering
        \includegraphics[width=\linewidth]{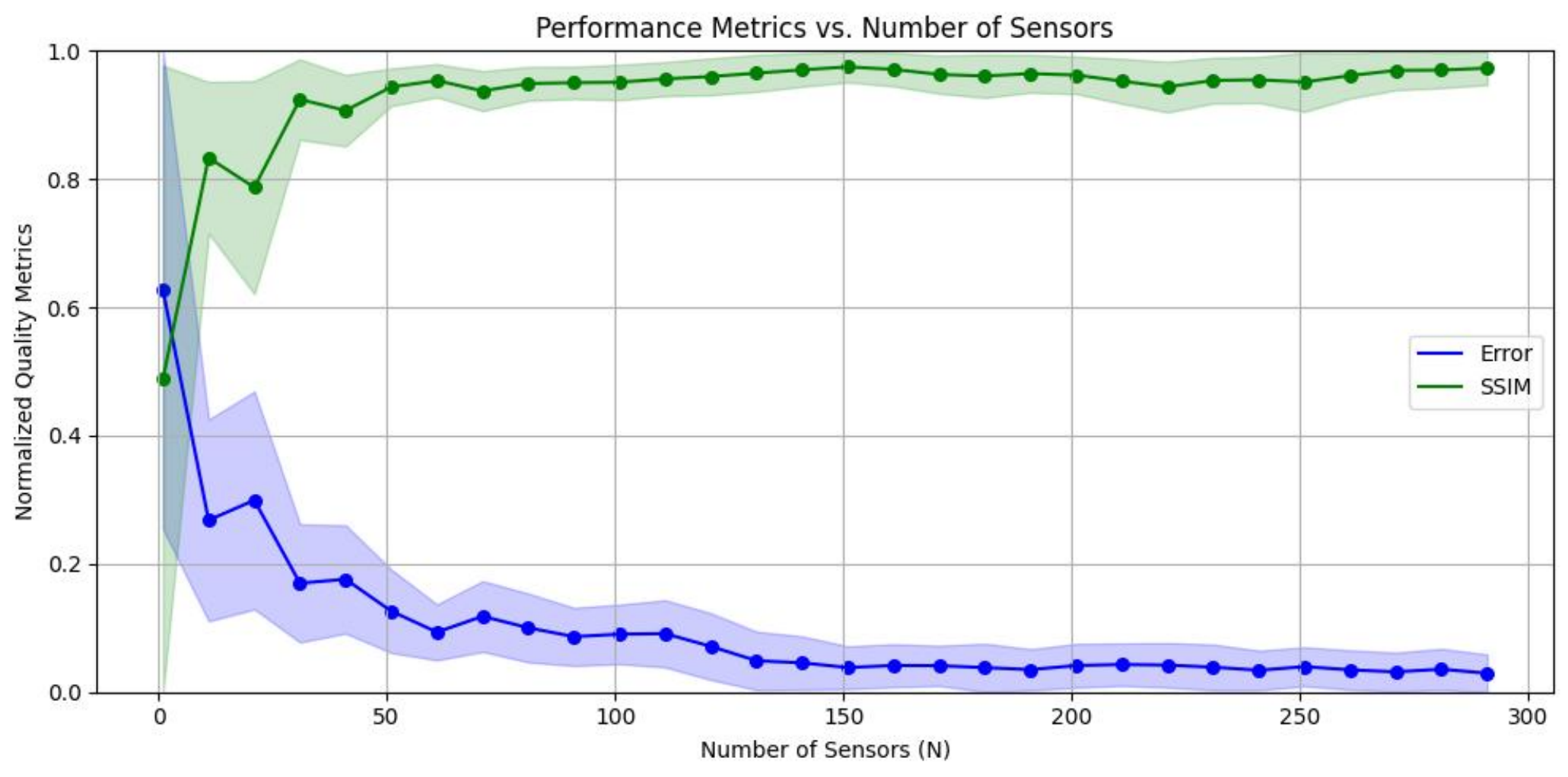}
        \caption{Relative Frobenius error and SSIM for pressure and oil saturation under QR-selected grid-wide placement.}
        \label{fig:metric_multifield_a}
    \end{subfigure}\hfill
    \begin{subfigure}[t]{0.49\textwidth}
        \centering
        \includegraphics[width=\linewidth]{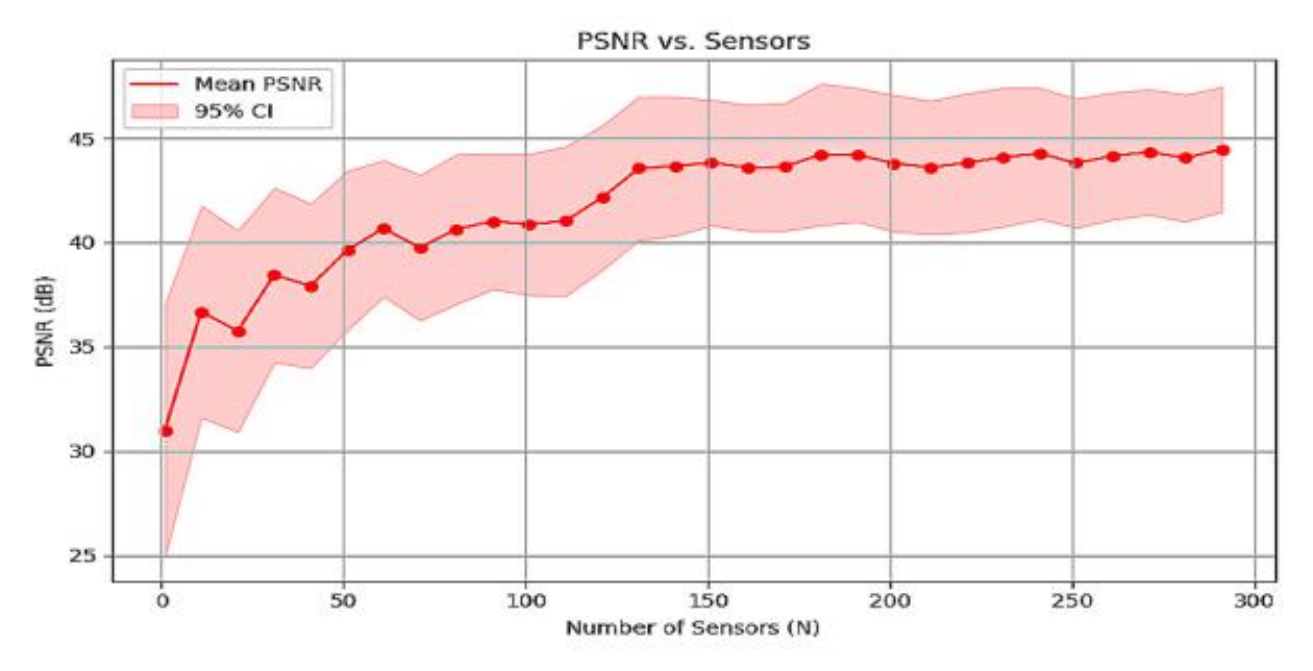}
        \caption{PSNR (dB) for pressure and oil saturation under QR-selected grid-wide placement.}
        \label{fig:metric_multifield_b}
    \end{subfigure}
    
    \caption{Multi-property (pressure + oil saturation) reconstruction metrics vs.\ number of QR-selected grid-wide sensor locations, all evaluated on the common training-fitted property-wise min--max normalized scale: (a) relative Frobenius error and SSIM; (b) PSNR (dB). These grid-wide budget sweeps are based on a QR ranking constructed with dictionary depth $W=480$; the same ranking later supports the largest evaluated configuration $N_{\mathrm{eval,max}}=299$ analyzed in Sec.~\ref{subsec:probabilistic_analysis}. This reported budget is a computational cutoff for the sweep, not the theoretical QR limit, and it satisfies $N_{\mathrm{eval,max}}\leq\min(W,|\Omega_{\mathrm{grid}}|)$. Mean curves aggregate the 10 control realizations with fixed geology; shaded bands indicate variability around those means (PSNR: 95\% CI). The joint tensor uses the fixed property order from Sec.~\ref{subsec:conventions_units}.}
    \label{fig:metric_multifield}
\end{figure}
\begin{figure}[htbp]
  \centering

  \begin{subfigure}[t]{0.32\textwidth}
    \centering
    \includegraphics[width=\linewidth]{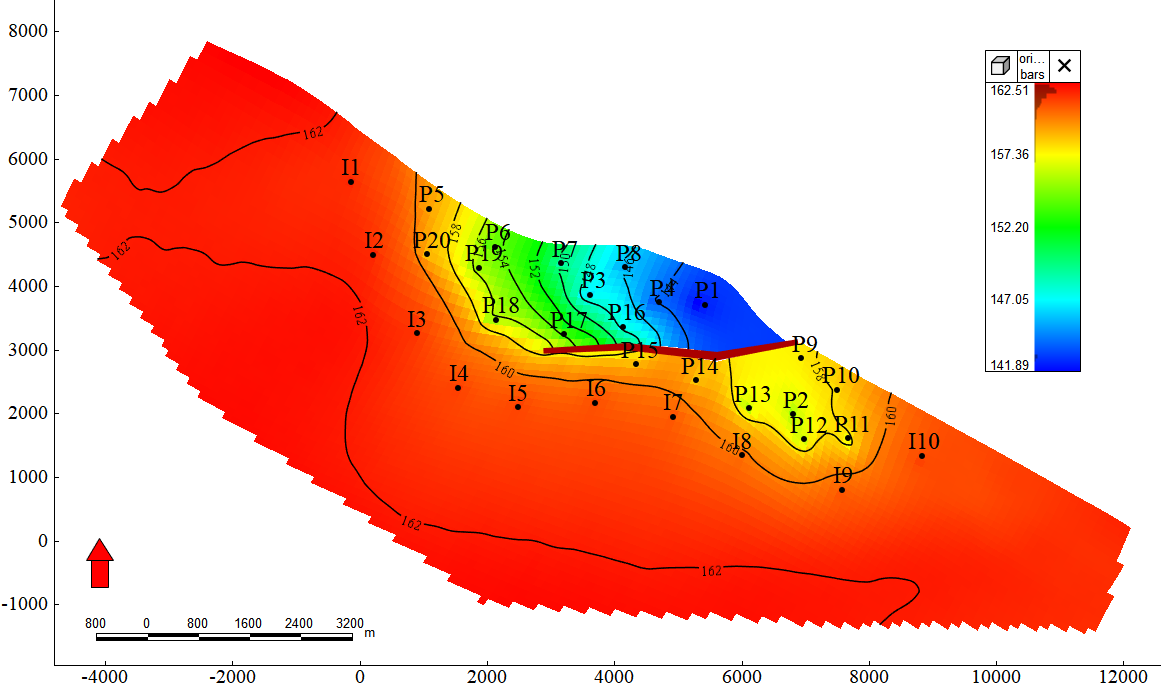}
    \caption{Ground truth (pressure).}
    \label{fig:tbmd_recon_gt_sensors_3d}
  \end{subfigure}\hfill
  \begin{subfigure}[t]{0.32\textwidth}
    \centering
    \includegraphics[width=\linewidth]{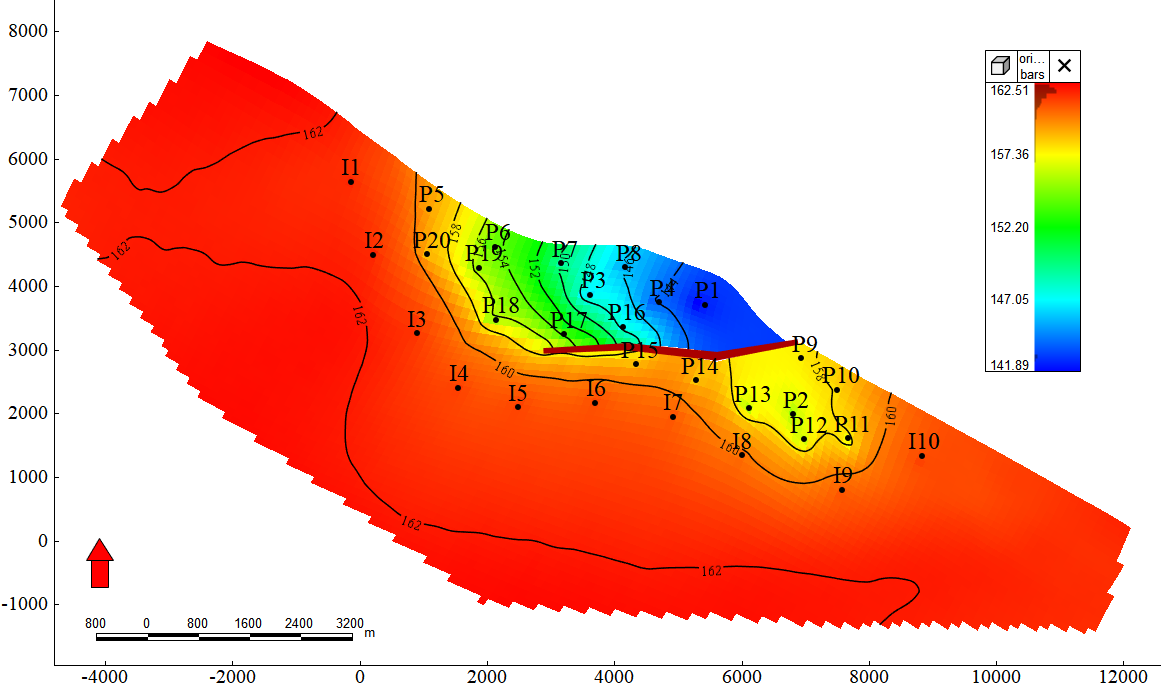}
    \caption{TBMD reconstruction for an illustrative 200-mode representative case; this basis size is used only for the displayed reconstruction example.}
    \label{fig:tbmd_recon_pred_sensors_3d}
  \end{subfigure}\hfill
  \begin{subfigure}[t]{0.32\textwidth}
    \centering
    \includegraphics[width=\linewidth]{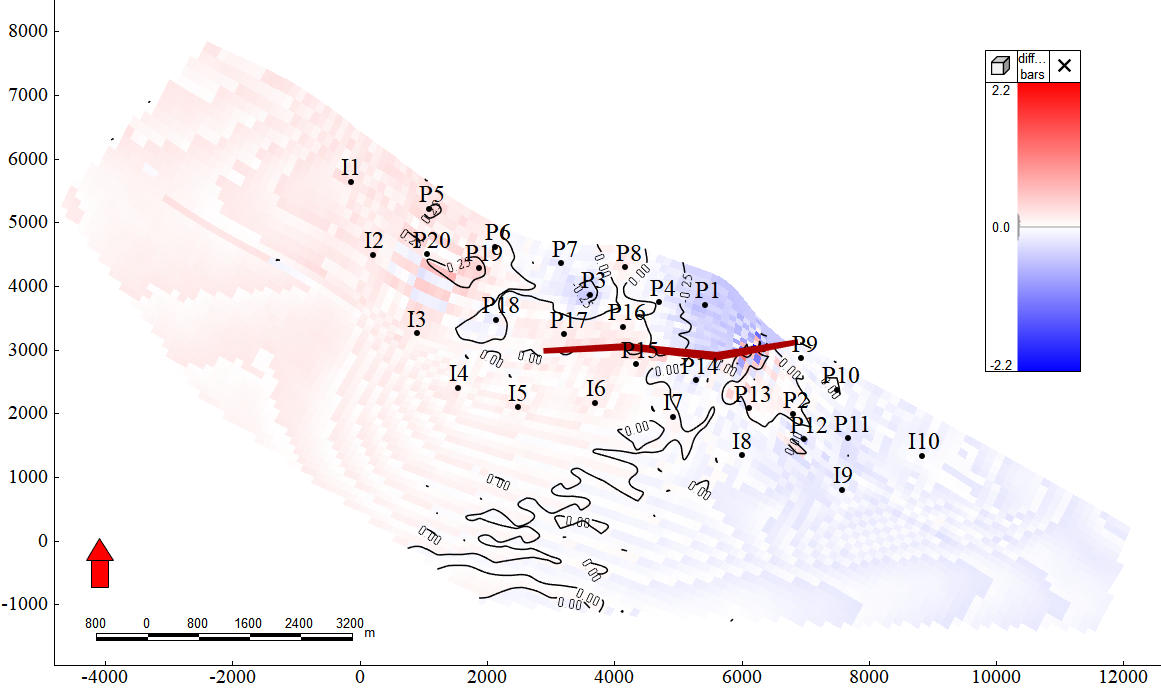}
    \caption{Per-cell signed error map (Recon--GT).}
    \label{fig:tbmd_recon_diff_sensors_3d}
  \end{subfigure}

  \caption{Representative pressure field reconstruction from QR-selected grid-wide sensor locations using TBMD+TBCS on one held-out run: (a) ground truth; (b) reconstruction; (c) per-cell signed error map (Recon--GT). Panel (b) shows an illustrative 200-mode representative case; this basis size is used only for the displayed reconstruction example and is not the dictionary depth used for the full grid-wide sensor-budget sweep.}
  \label{fig:tbmd_sensor_recon_sensors_3d}
\end{figure}

For the grid-wide sensor-budget sweeps reported in Figures~\ref{fig:metric_convergence} and~\ref{fig:metric_multifield}, the QR ranking is constructed with dictionary depth $W=480$. The largest evaluated configuration used later for the cluster diagnostics is $N_{\mathrm{eval,max}}=299$, which is a computational cutoff for the sweep rather than the theoretical QR limit and satisfies $N_{\mathrm{eval,max}}\leq\min(W,|\Omega_{\mathrm{grid}}|)$. The representative map-based reconstructions shown later in Figures~\ref{fig:tbmd_sensor_recon_sensors_3d} and~\ref{fig:tbmd_sensor_recon_sensors_4d} instead use a 200-mode dictionary only for illustrative visualization and qualitative comparison. These two choices serve different purposes and should not be conflated.
\par

\FloatBarrier
\FloatBarrier

\FloatBarrier
\FloatBarrier

Figure~\ref{fig:tbmd_sensor_recon_sensors_3d} presents a representative visualization of TBMD pressure field reconstruction from QR-selected grid-wide sensor locations on one held-out run. The following map is an illustrative reconstruction example using a 200-mode basis. This value is not used as a universal dictionary depth for all grid-wide sensor-budget sweeps. The left panel shows the ground-truth pressure slice, the center displays the reconstructed field, and the right panel shows the per-cell signed error map, computed as reconstruction minus ground truth (Recon--GT). The visual comparison indicates high reconstruction accuracy, with only minor localized discrepancies observable in the signed difference map.
\par

\FloatBarrier
\FloatBarrier

\subsubsection{Multi-property TBMD Reconstruction (Pressure + Oil Saturation)}
\label{subsubsec:full_coverage_multi}
\par
Beyond the single-property analysis, we evaluate TBMD on a joint \emph{multi-property} tensor that simultaneously includes pressure and oil saturation. For each run, the data are arranged as
$\mathcal{X}^{(r)}\in\mathbb{R}^{139 \times 48 \times 2 \times 134}$,
where the third mode follows the fixed property order from Sec.~\ref{subsec:conventions_units}. 
This formulation allows the extraction of shared spatiotemporal modes across both properties, 
potentially improving compression efficiency and structural coherence.
\par
In the 4D case considered here, $\mathcal{X}\in\mathbb{R}^{I\times J\times 2\times L}$, 
the third mode is a categorical index of reservoir properties. 
To avoid degeneracy of this categorical mode, we fix its rank to the full size, $r_{\mathrm{prop}}=2$, 
and select minimal ranks for the remaining modes $(r_x,r_y,r_t)$ to satisfy a target-accuracy criterion:
\[
\frac{\|\mathcal{X} - \hat{\mathcal{X}}(r_x, r_y, r_{\mathrm{prop}}{=}2, r_t)\|_F}
     {\|\mathcal{X}\|_F}
\le \varepsilon
\quad \text{or} \quad
E_{\mathrm{retained}}(r_x, r_y, r_t) \ge \eta,
\]
where $\varepsilon$ is the relative-error tolerance and $E_{\mathrm{retained}}$ denotes the retained energy fraction. 
This ensures a joint reduced-order dictionary that captures the coupled dynamics of pressure and oil saturation 
without collapsing the property mode.
\par
As in Sec.~\ref{subsubsec:full_coverage_single}, we use an $80/20$ temporal split:
$80\%$ of time steps (per run) for mode extraction (training) and $20\%$ for evaluation.
The TBMD factors and core are fitted strictly on the training portion; reconstruction
metrics (Frobenius error, SSIM, PSNR) are computed on the held-out $20\%$, per property.
Before assembling the joint 4D tensor used by TBMD, QR, and TBCS, pressure and oil saturation are scaled separately with the training-fitted affine maps from Sec.~\ref{subsec:conventions_units} and stacked using that fixed property order; the joint residual therefore compares both channels on the same dimensionless scale rather than in mixed physical units.
\par
Figure~\ref{fig:metric_multifield} summarizes the reconstruction performance over the full grid 
for both pressure and oil saturation. 
As in Figure~\ref{fig:metric_convergence}, the plotted mean curves aggregate held-out metrics from the sequential 80/20 split over all 10 control realizations with fixed geology.
Overall, errors remain low across both properties; pressure is typically reconstructed slightly better 
due to its smoother spatial structure, whereas oil saturation exhibits sharper fronts. 
The SSIM and PSNR trends confirm that structural information is well preserved in both channels, 
indicating that joint-mode decomposition is feasible and effective for multi-property reservoir models.
\par
\begin{figure}[htbp]
  \centering

  \begin{subfigure}[t]{0.32\textwidth}
    \centering
    \includegraphics[width=\linewidth]{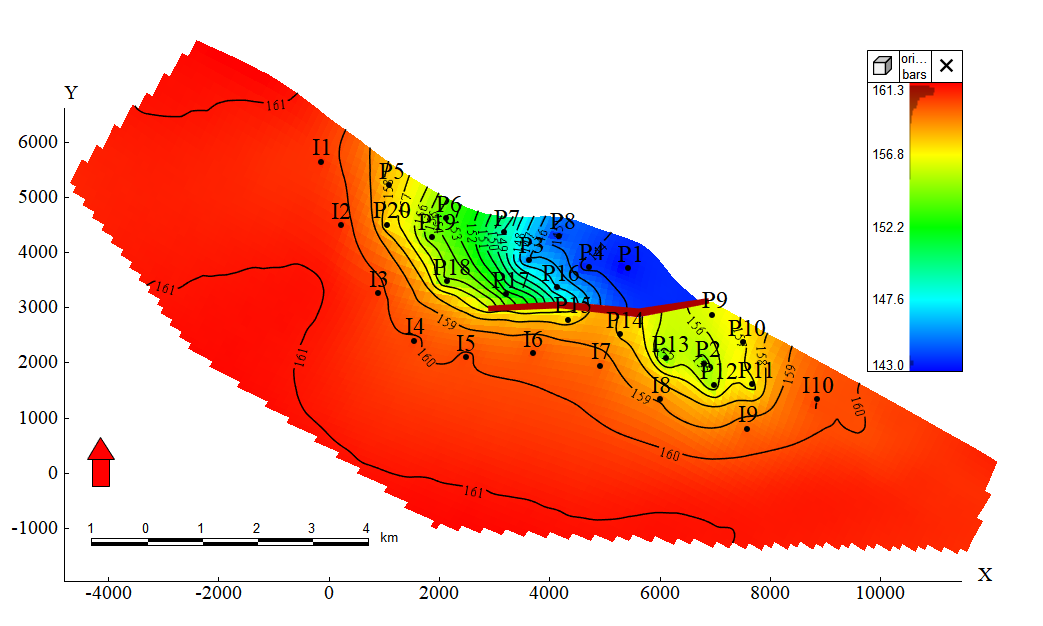}
    \caption{Ground truth (pressure).}
    \label{fig:tbmd_recon_gt_sensors_4d}
  \end{subfigure}\hfill
  \begin{subfigure}[t]{0.32\textwidth}
    \centering
    \includegraphics[width=\linewidth]{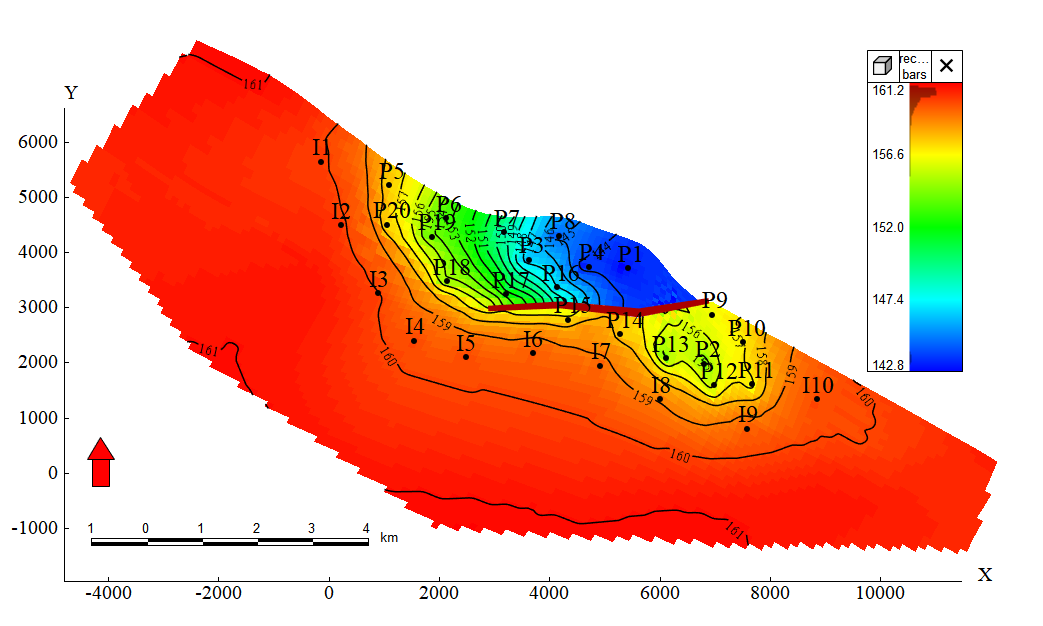}
    \caption{TBMD reconstruction for an illustrative 200-mode representative case; this basis size is used only for the displayed reconstruction example.}
    \label{fig:tbmd_recon_pred_sensors_4d}
  \end{subfigure}\hfill
  \begin{subfigure}[t]{0.32\textwidth}
    \centering
    \includegraphics[width=\linewidth]{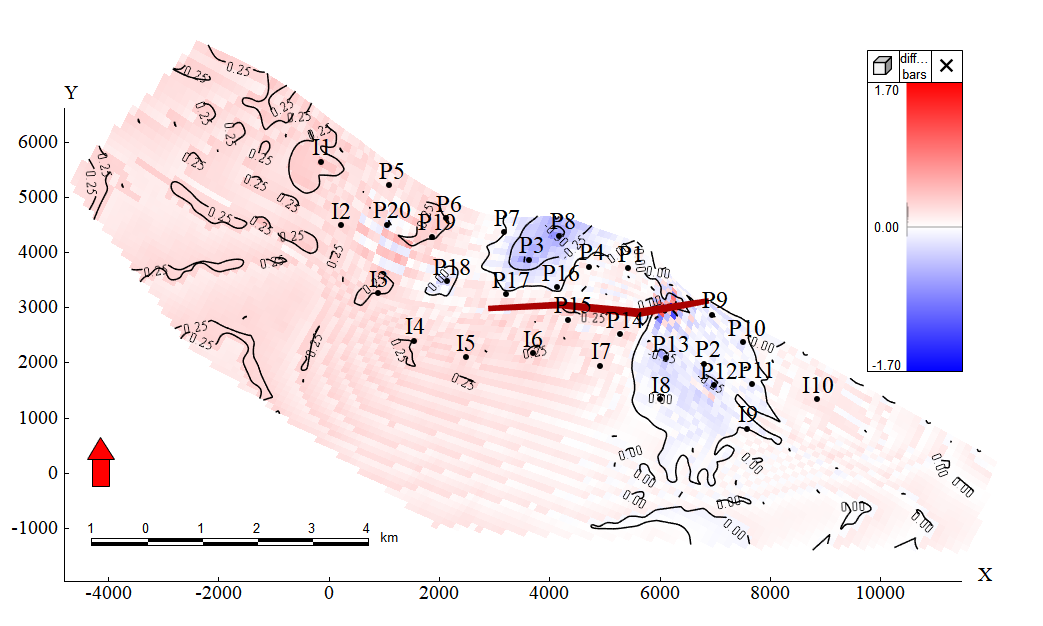}
    \caption{Per-cell signed error map (Recon--GT).}
    \label{fig:tbmd_recon_diff_sensors_4d}
  \end{subfigure}

  \caption{Representative pressure field reconstruction from the 4D TBMD representation using QR-selected grid-wide sensor locations on one held-out run.
  Each panel shows a 2D slice from a single-run tensor $\mathcal{X}^{(r)}\in\mathbb{R}^{139\times48\times2\times134}$: (a) ground-truth pressure; (b) TBMD reconstruction for an illustrative 200-mode representative case; (c) per-cell signed error map (Recon--GT). This basis size is used only for the displayed reconstruction example and is not the dictionary depth used for the full grid-wide sensor-budget sweep.}
  \label{fig:tbmd_sensor_recon_sensors_4d}
\end{figure}

A visual comparison of reconstruction results derived from the 4D TBMD representation 
from QR-selected grid-wide sensor locations is presented in Figure~\ref{fig:tbmd_sensor_recon_sensors_4d}. 
The maps illustrate the quality of pressure field recovery, demonstrating that the 4D formulation 
successfully captures the main spatiotemporal patterns while maintaining consistency between the two physical properties. As in Figure~\ref{fig:tbmd_sensor_recon_sensors_3d}, the field map in Figure~\ref{fig:tbmd_sensor_recon_sensors_4d} uses a representative 200-mode basis only for qualitative visualization; the full grid-wide budget sweeps and the later clustering analysis use dictionary depth $W=480$.
\par

\FloatBarrier
\FloatBarrier

These results indicate that TBMD can represent the dominant dynamics 
of multiple reservoir parameters within a unified 4D tensor representation. 
Joint modeling of pressure and oil saturation can be useful when their coupled behavior 
is to be represented by a shared reduced-order model.

\subsection{Field Reconstruction from Existing-Well Measurements}
\label{subsec:well_based_measurements}
\subsubsection{Single-property TBMD Reconstruction from Existing-Well Measurements (Pressure)}
\label{subsubsec:well_based_single}
\par
The Brugge benchmark model used for this experiment contains 30 wells (20 producers and 10 injectors)~\cite{peters2010Brugge, TNO2009}. Accordingly, the well-only admissible set is the physical well set $\mathcal{W}$ with $|\mathcal{W}|=30$. The pressure field is defined on \(139\times48=6672\) grid cells, but measurements are available only at the 30 well parent cells (about \(0.45\%\) of the grid). At each held-out time step, a selected subset $\mathcal{W}_N\subseteq\mathcal{W}$ provides $N$ scalar pressure measurements, one per physical well. The TBMD basis is fitted separately on the first \(80\%\) of time steps of each run and evaluated on the corresponding held-out \(20\%\).
\par
\begin{figure}[htbp]
    \centering
    \begin{subfigure}[t]{0.49\textwidth}
        \centering
        \includegraphics[width=\linewidth]{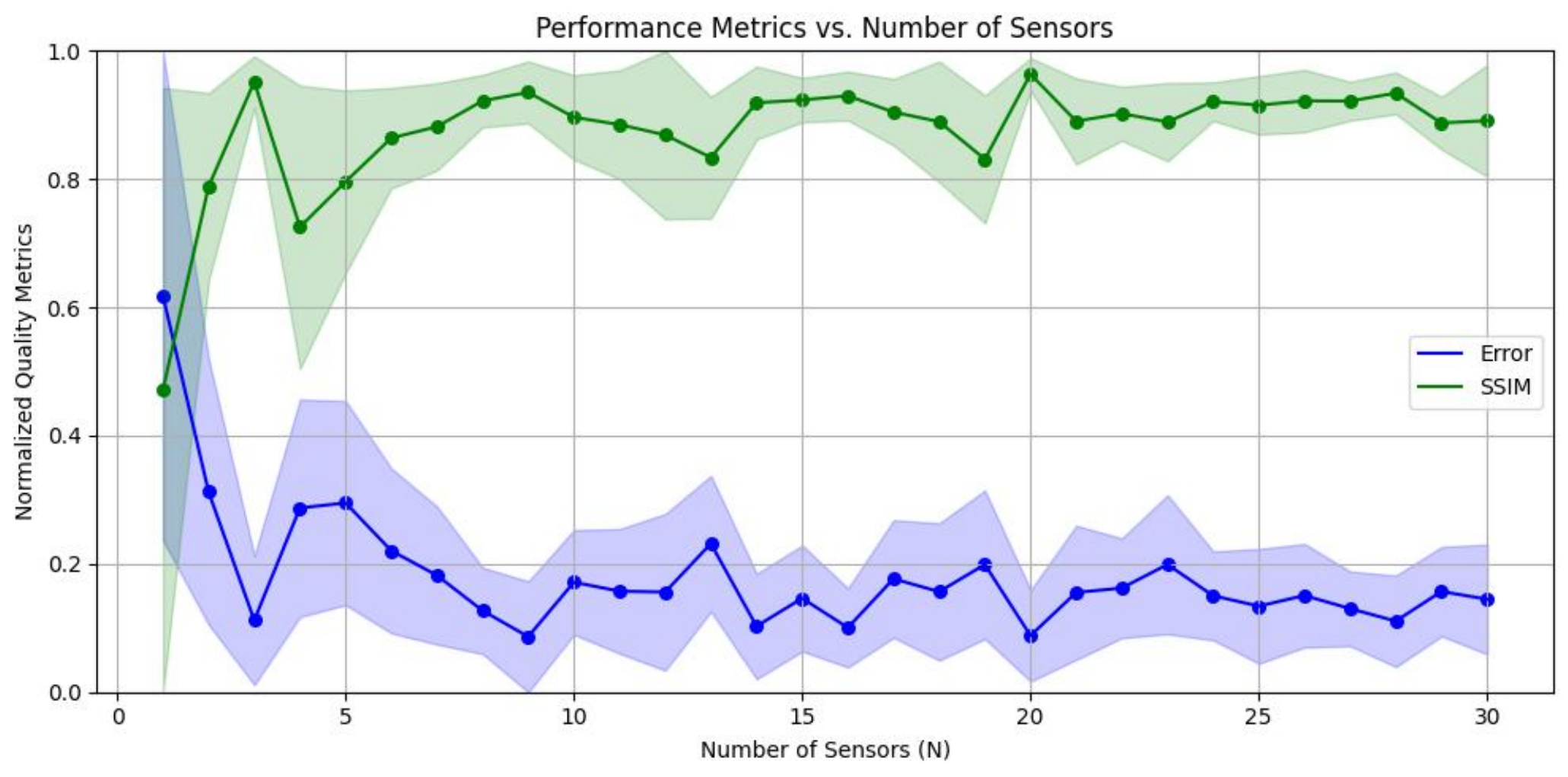}
        \caption{Relative Frobenius error and SSIM vs.\ number of instrumented wells in the random well-subset test.}
        \label{fig:metric_convergence_wells_a}
    \end{subfigure}\hfill
    \begin{subfigure}[t]{0.49\textwidth}
        \centering
        \includegraphics[width=\linewidth]{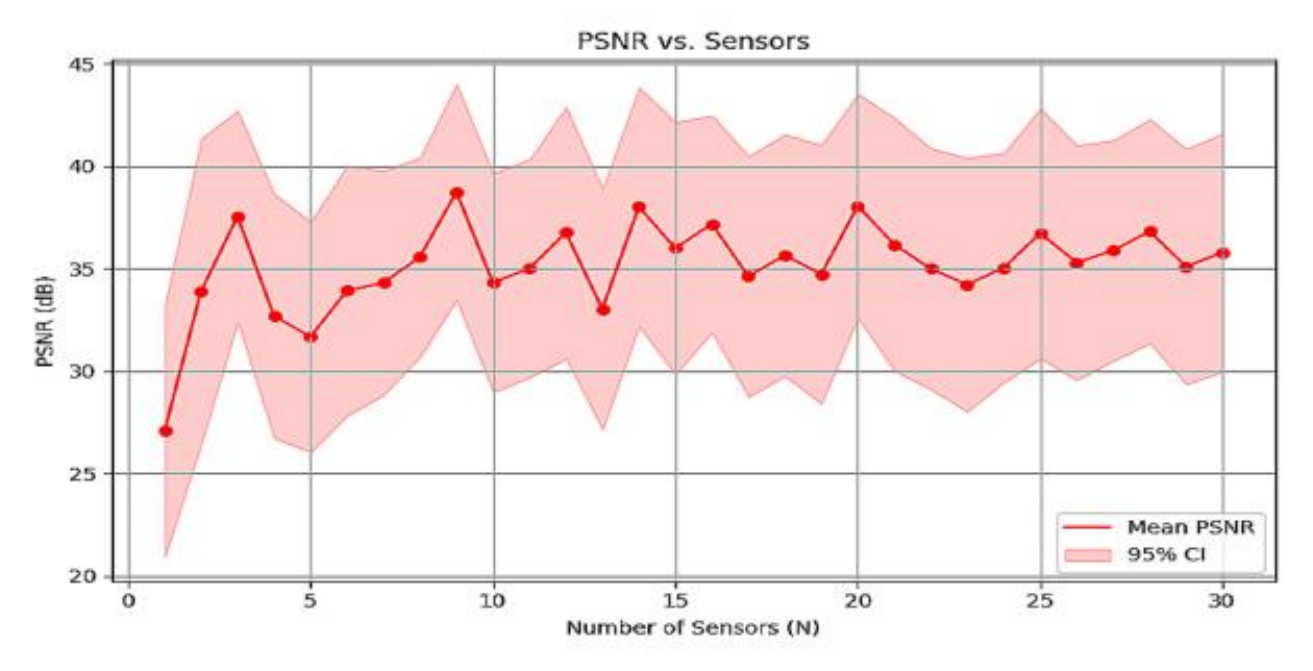}
        \caption{PSNR (dB) vs.\ number of instrumented wells in the random well-subset test.}
        \label{fig:metric_convergence_wells_b}
    \end{subfigure}
    \caption{Pressure field reconstruction metrics in the well-only random-subset test vs.\ number of instrumented wells, all evaluated on the common training-fitted property-wise min--max normalized scale: (a) relative Frobenius error and SSIM; (b) PSNR (dB). The admissible set is restricted to the 30 existing physical wells, $|\mathcal{W}|=30$, and the budget $N$ counts selected wells rather than property-specific well channels. Mean curves aggregate the 10 well-control scenarios and the random well subsets; shaded bands indicate variability around those means (PSNR: 95\% CI). The curves are not expected to be strictly monotonic because each budget is evaluated as a random well-subset feasibility test, and the \(N=30\) case corresponds to the fixed all-well configuration rather than a nested ranked selection.}
    \label{fig:metric_convergence_wells}
\end{figure}

For each sensor budget \(N\), we form random subsets of \(N\) wells from the 30 available wells \emph{without replacement}, i.e., the same well cannot be chosen twice within one subset. In contrast to the grid-wide setting, the well-only experiments report the sensor budget in terms of physical wells. Although multiple physical properties may be represented in the tensor, the well-only admissible set is not counted as \(30\times2\) property-specific channels. It is counted as 30 available physical wells:
\[
|\mathcal{W}|=30.
\]
Thus, the maximum well-only budget is \(N=30\), provided that the dictionary depth satisfies \(W\geq30\). Figure~\ref{fig:metric_convergence_wells} therefore reports a \emph{well-only feasibility/stress test} under existing-well constraints rather than a QR-ranked selection within the admissible well set. The mean curves and variability bands are obtained by averaging over the 10 held-out runs, the held-out test snapshots within each run, and the random well subsets associated with each budget.
\par
Figure~\ref{fig:metric_convergence_wells} shows that reconstruction quality improves overall as the well budget grows from very small to intermediate values, but the reported curve is not strictly monotonic because the well-only experiment is a random-subset feasibility test under fixed existing-well geometry rather than a nested QR-ranked well-selection procedure. Increasing the budget from \(N=1\) to \(N=10\) substantially reduces the relative Frobenius error and improves SSIM and PSNR. Around \(N=20\), the reported typical values are approximately error \(\approx 0.09\), SSIM \(\approx 0.96\), and PSNR \(\approx 38\) dB.
\par
For the full 30-well configuration, all existing wells are included. In this case, the mean metrics are approximately error \(\approx 0.15\), SSIM \(\approx 0.89\), and PSNR \(\approx 35.8\) dB. The slight degradation relative to the best intermediate budgets indicates that adding all available wells does not necessarily produce a monotonic improvement under the fixed TBCS regularization and existing-well geometry. Some wells may provide spatially redundant or less informative measurements, and their inclusion can alter the recovered modal coefficients in a way that slightly worsens aggregate held-out metrics. Therefore, the well-only curves should be interpreted as a feasibility study and empirical stability check under operational constraints, not as a monotone sensor-ranking result.
\par

\FloatBarrier
\FloatBarrier

\FloatBarrier
\begin{figure}[htbp]
    \centering
    \begin{subfigure}[t]{0.49\textwidth}
        \centering
        \includegraphics[width=\linewidth]{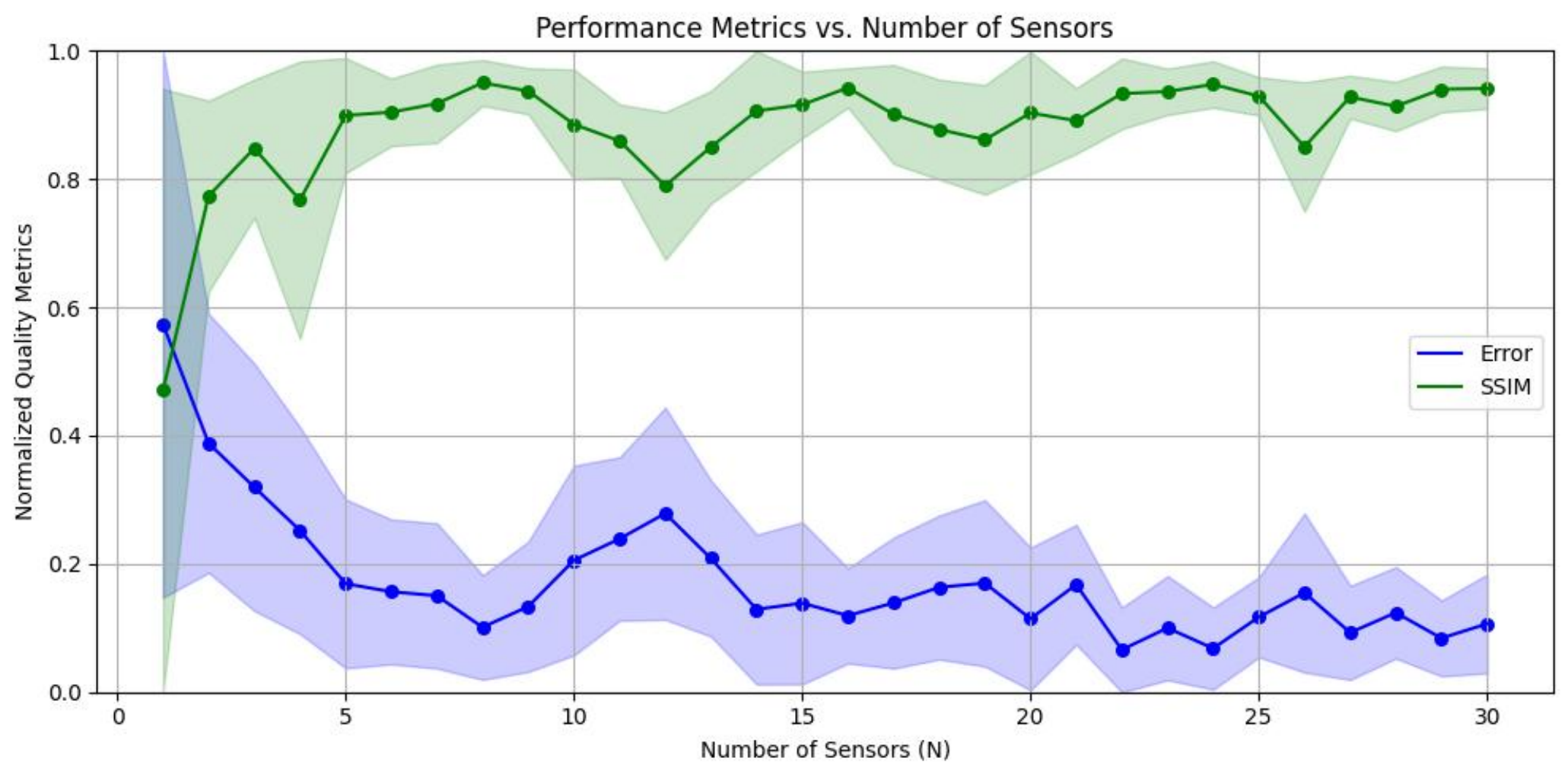}
        \caption{Relative Frobenius error and SSIM for pressure and oil saturation.}
        \label{fig:metric_multifield_wells_a}
    \end{subfigure}\hfill
    \begin{subfigure}[t]{0.49\textwidth}
        \centering
        \includegraphics[width=\linewidth]{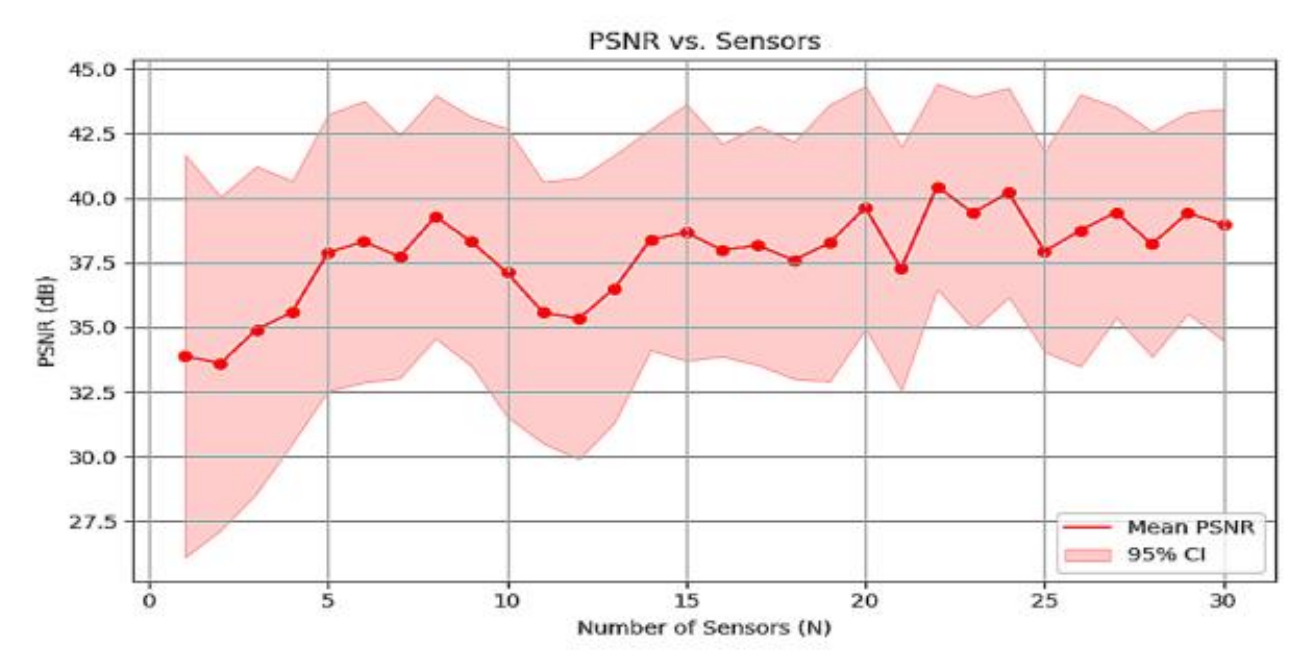}
        \caption{PSNR (dB) for pressure and oil saturation.}
        \label{fig:metric_multifield_wells_b}
    \end{subfigure}
    \caption{Joint (pressure + oil saturation) reconstruction metrics in the well-only random-subset test vs.\ number of instrumented wells, all evaluated on the common training-fitted property-wise min--max normalized scale: (a) relative Frobenius error and SSIM; (b) PSNR (dB). The admissible set is restricted to the 30 existing physical wells, $|\mathcal{W}|=30$, and the budget $N$ counts selected wells rather than property-specific well channels. Mean curves aggregate the 10 well-control scenarios and the random well subsets; shaded bands indicate variability around those means (PSNR: 95\% CI). The joint tensor uses the fixed property order from Sec.~\ref{subsec:conventions_units}.}
    \label{fig:metric_multifield_wells}
\end{figure}
\FloatBarrier

\FloatBarrier
\begin{figure}[htbp]
  \centering

  \begin{subfigure}[t]{0.32\textwidth}
    \centering
    \includegraphics[width=\linewidth]{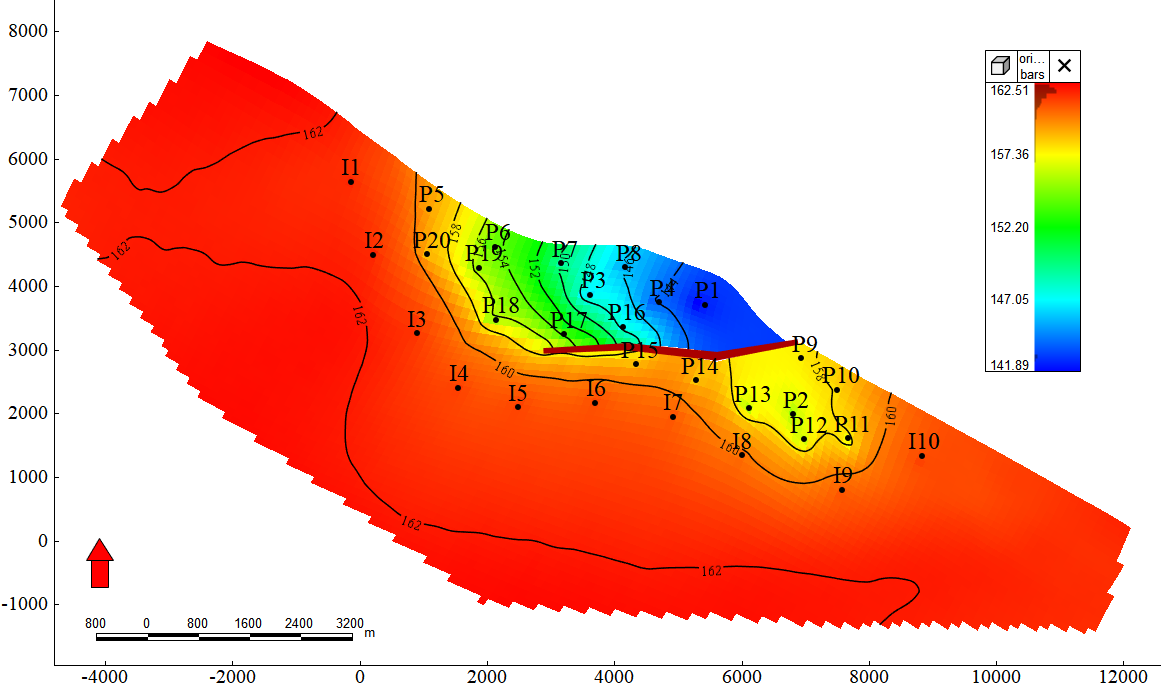}
    \caption{Ground truth (pressure).}
    \label{fig:tbmd_recon_gt_wells_3d}
  \end{subfigure}\hfill
  \begin{subfigure}[t]{0.32\textwidth}
    \centering
    \includegraphics[width=\linewidth]{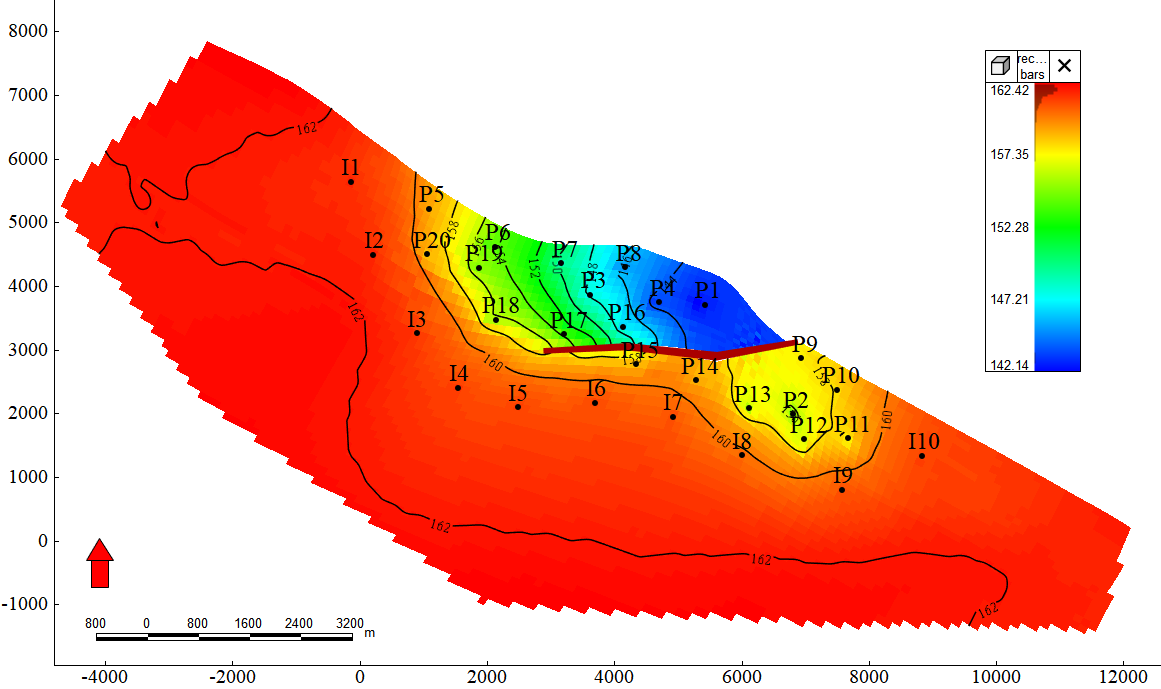}
    \caption{TBMD reconstruction (all 30 physical wells instrumented).}
    \label{fig:tbmd_recon_pred_wells_3d}
  \end{subfigure}\hfill
  \begin{subfigure}[t]{0.32\textwidth}
    \centering
    \includegraphics[width=\linewidth]{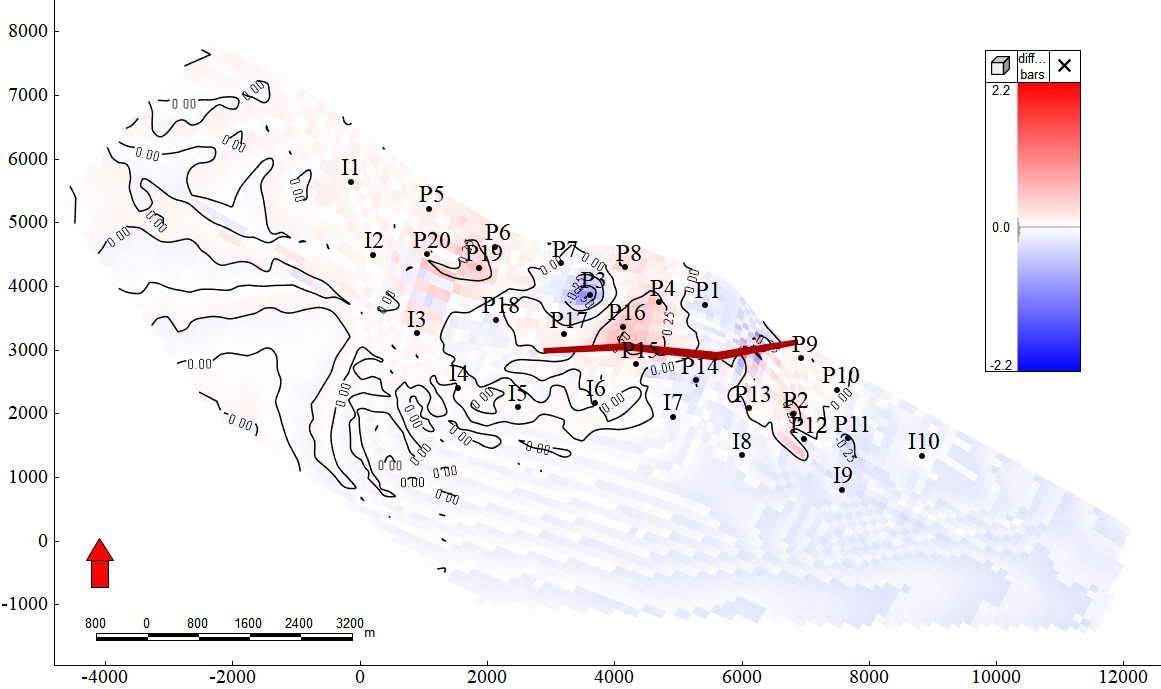}
    \caption{Per-cell signed error map (Recon--GT).}
    \label{fig:tbmd_recon_wells_diff_3d}
  \end{subfigure}

  \caption{Representative reconstruction of the pressure field from existing-well measurements using TBMD+TBCS with all 30 physical wells instrumented on one held-out run: (a) ground truth; (b) reconstructed field; (c) per-cell signed error map (Recon--GT). Here the admissible set is restricted to $|\mathcal{W}|=30$, and the budget $N$ counts wells rather than property-specific channels.}
  \label{fig:tbmd_recon_wells_3d}
\end{figure}
\FloatBarrier

Figure~\ref{fig:tbmd_recon_wells_3d} illustrates the all-well pressure reconstruction used for this qualitative check. The all-well case remains qualitatively interpretable and captures the dominant pressure-gradient structure, but it is not the best-performing budget in this single-property well-only experiment. This non-monotonicity is consistent with the fact that the well-only setup uses fixed existing-well locations rather than nested QR-ranked sensor placement.

\subsubsection{Multi-property TBMD Reconstruction from Existing-Well Measurements (Pressure + Oil Saturation)}
\label{subsubsec:well_based_multi}
\par
We evaluate joint reconstruction of pressure and oil saturation using the fixed property order from Sec.~\ref{subsec:conventions_units}, with the same well-only constraint and the same random-subset protocol as above: measurements are available only at the same 30 existing physical wells, i.e., on the admissible set $\mathcal{W}$ with $|\mathcal{W}|=30$ (from a full grid of \(6672\) cells). At each held-out time step, a selected subset $\mathcal{W}_N\subseteq\mathcal{W}$ contributes two scalar measurements per well in the fixed property order. Here again, the budget $N$ counts wells rather than well--property channels.
\par
Figure~\ref{fig:metric_multifield_wells} provides direct quantitative evidence that the joint TBMD+TBCS model remains effective under sparse instrumentation in this well-only feasibility test. From the mean curves, increasing the number of instrumented wells from \(N=1\) to \(N=10\) reduces the relative error from about \(0.57\) to \(0.20\) (about \(65\%\) reduction), increases SSIM from about \(0.47\) to \(0.88\), and raises PSNR from about \(33.8\,\mathrm{dB}\) to \(37\,\mathrm{dB}\).
\par
At \(N=20\), the reported metrics are error \(\approx 0.11\), SSIM \(\approx 0.90\), and PSNR \(\approx 39.5\,\mathrm{dB}\). At \(N=30\), the mean metrics are error \(\approx 0.10\), SSIM \(\approx 0.94\), and PSNR \(\approx 39\,\mathrm{dB}\), indicating internally consistent joint reconstruction under the well-only feasibility protocol.
\par
\begin{figure}[htbp]
  \centering

  \begin{subfigure}[t]{0.32\textwidth}
    \centering
    \includegraphics[width=\linewidth]{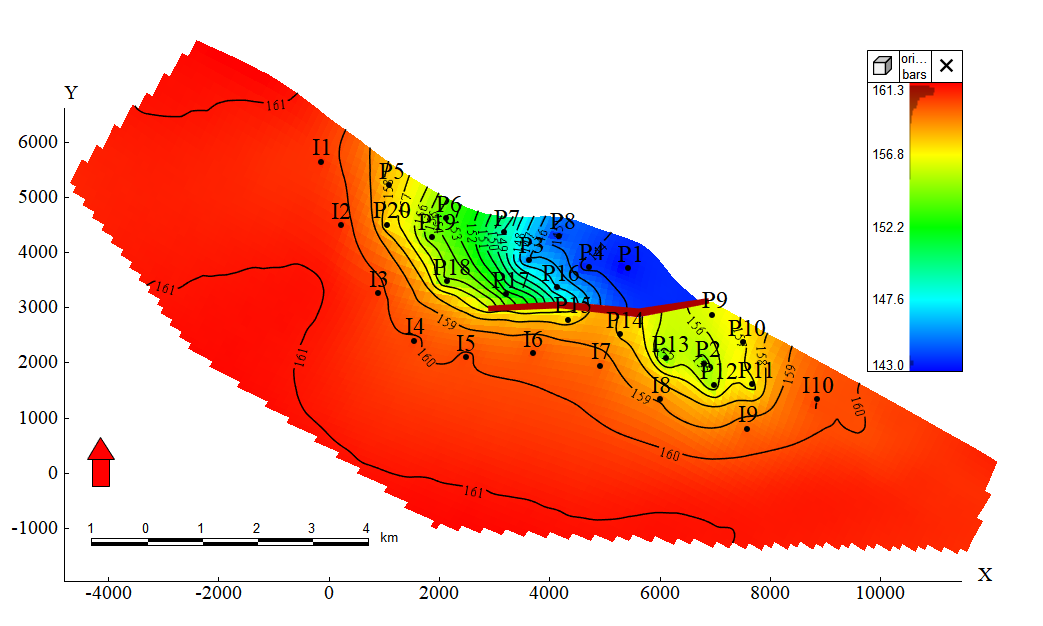}
    \caption{Ground truth (pressure).}
    \label{fig:tbmd_gt_wells_4d}
  \end{subfigure}\hfill
  \begin{subfigure}[t]{0.32\textwidth}
    \centering
    \includegraphics[width=\linewidth]{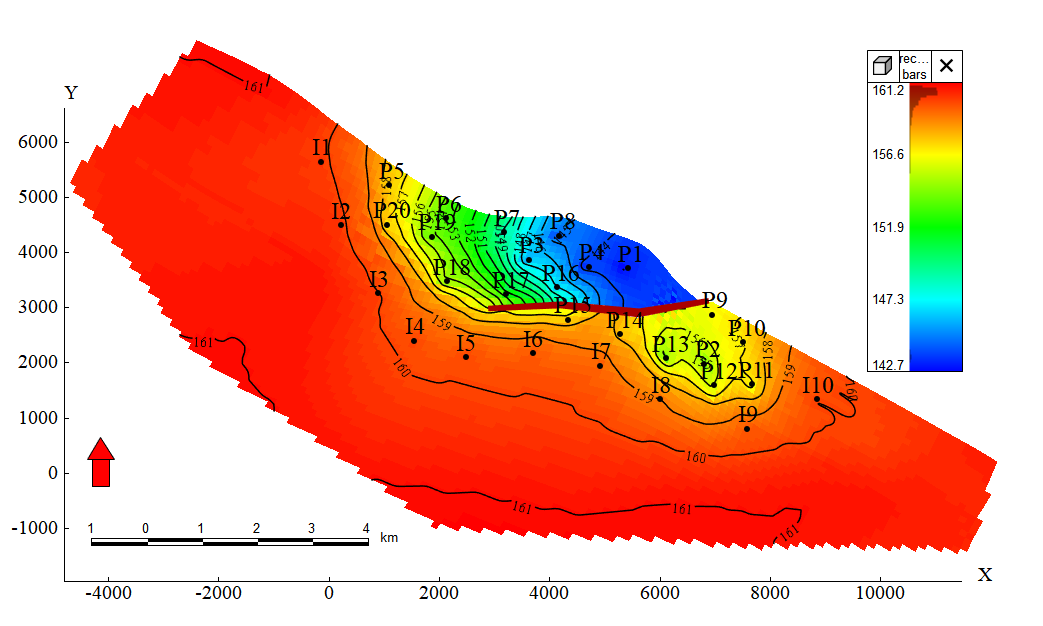}
    \caption{4D TBMD reconstruction (pressure; all 30 physical wells instrumented).}
    \label{fig:tbmd_rec_wells_4d}
  \end{subfigure}\hfill
  \begin{subfigure}[t]{0.32\textwidth}
    \centering
    \includegraphics[width=\linewidth]{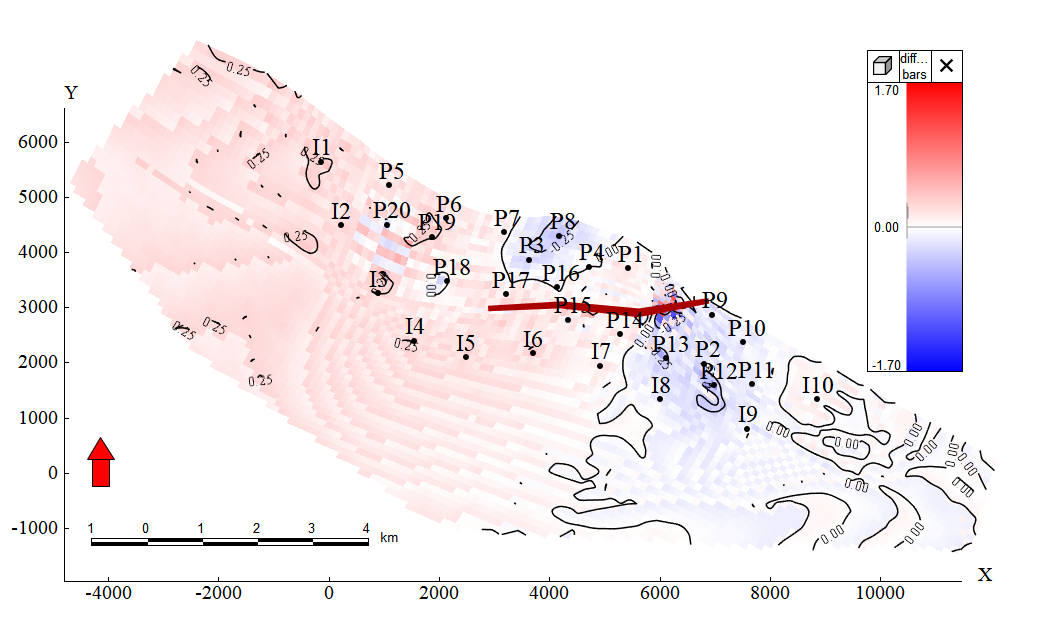}
    \caption{Per-cell signed error map (Recon--GT).}
    \label{fig:tbmd_recon_diff_wells_4d}
  \end{subfigure}

  \caption{Representative pressure field reconstruction from the 4D TBMD representation using \emph{well-only} measurements with all 30 existing physical wells instrumented on one held-out run.
  Each panel shows a 2D slice from a single-run tensor $\mathcal{X}^{(r)}\in\mathbb{R}^{139\times48\times2\times134}$: (a) ground-truth pressure; (b) reconstructed field; (c) per-cell signed error map (Recon--GT). In this well-only setting, $|\mathcal{W}|=30$ and $N$ counts physical wells rather than property-specific channels.}
  \label{fig:tbmd_recon_wells_4d}
\end{figure}

The recoverability claim is also supported visually by Figure~\ref{fig:tbmd_recon_wells_4d}: for the 30-well case, the reconstructed pressure map reproduces the large-scale gradient and injector--producer connectivity, while the largest residuals are localized in peripheral zones with weaker measurement support.
\par

\FloatBarrier
\FloatBarrier

These quantitative and visual results support the hypothesis of algorithm effectiveness: a shared 4D TBMD basis enables accurate multi-property recovery from sparse well-only data in the existing-well feasibility test, with most quality gains achieved already in the \(N\approx10\text{--}20\) range.

\subsection{Probabilistic Analysis of Grid-wide TBMD--QR Sensor Distribution}
\label{subsec:probabilistic_analysis}
\par
This subsection analyzes the \emph{grid-wide} QR ranking rather than the random well-subset curves of Sec.~\ref{subsec:well_based_measurements}. Within the TBMD+QR+TBCS framework, tensor QR factorization identifies informative spatial--property locations by pivoting columns of the unfolded dictionary $\mathcal{A}_{(4)}$ (equivalently, mode-4 fibers of the dictionary tensor, i.e., modal-amplitude vectors rather than physical time traces). In this part of the study, the admissible set is $\Omega=\Omega_{\mathrm{grid}}$, where $\Omega_{\mathrm{grid}}$ contains admissible spatial--property locations on the simulation grid. For the grid-wide sensor-budget sweep and the subsequent clustering analysis of QR-selected locations, the QR ranking was computed with dictionary depth $W=480$. The algorithmic admissibility condition is therefore
\[
N \leq \min(W,|\Omega_{\mathrm{grid}}|).
\]
In this analysis, we did not evaluate the full admissible range up to $W$. Instead, for computational efficiency, we restricted the diagnostic sweep to the largest evaluated budget
\[
N_{\mathrm{eval,max}}=299.
\]
This value is a numerical cutoff for the reported sweep, not the theoretical maximum number of QR pivots. It satisfies the admissibility condition because
\[
N_{\mathrm{eval,max}}=299 < W=480,
\]
and the grid-wide admissible mask contains more than 299 spatial--property locations. The representative reconstruction maps shown elsewhere use a 200-mode basis only for illustrative visualization and do not define the dictionary depth used in this QR-distribution analysis. The purpose of this sweep is to quantify the expected improvement in reconstruction quality as the number of selected grid-wide sensors increases.
\par
\begin{figure}[htbp]
  \centering
  \includegraphics[width=0.85\textwidth, height=0.6\textheight, keepaspectratio]{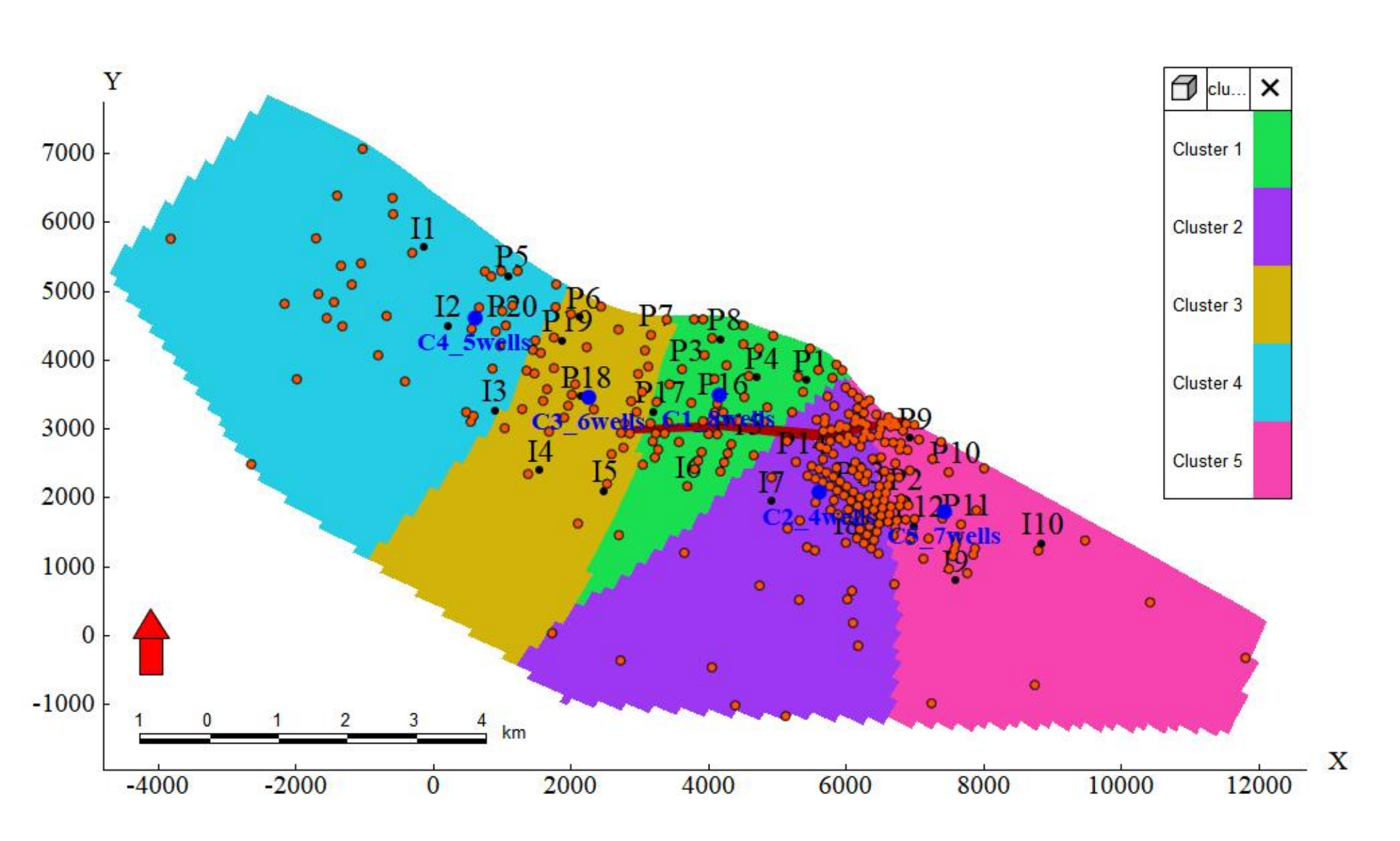}
  \caption{Cluster map for the largest evaluated grid-wide TBMD--QR configuration. Existing wells are blue, QR-selected grid cells are red, and shaded regions denote clusters \(C1\)--\(C5\).}
  \label{fig:cluster_map}
\end{figure}

To evaluate the spatial distribution of TBMD--QR sensors, we define a reproducible 2D cluster partition on the admissible spatial grid obtained by projecting $\Omega_{\mathrm{grid}}$ onto physical cells. Let $\Omega_{\mathrm{grid}}^{xy}\subseteq\{1,\dots,I\}\times\{1,\dots,J\}$ denote this spatial projection, and let $\{p_w\}_{w\in\mathcal{W}}\subseteq\mathbb{R}^2$ be the 2D coordinates of the 30 well parent-cell centroids. The clustering step is performed only on the well coordinates $\{p_w\}_{w\in\mathcal{W}}$: we apply k-means to them and select the reported number of clusters via the silhouette coefficient, with confirmation by the Gap statistic; both criteria indicate \(k=5\). The fitted centroids $\{\mu_k\}_{k=1}^5$ then induce a full-grid partition by nearest-centroid assignment,
\[
\ell(c)\;=\;\arg\min_{k\in\{1,\dots,5\}} \norm{p(c)-\mu_k}_2,
\qquad
C_k\;=\;\{c\in\Omega_{\mathrm{grid}}^{xy}:\ell(c)=k\},
\]
where $p(c)$ is the 2D centroid of cell $c$. Thus, the well-based k-means stage induces a Voronoi-type partition of the full admissible grid, and this induced partition is the one used for Table~\ref{tab:cluster_probabilities}, Figure~\ref{fig:cluster_map}, and all cell-level fractions below. Accordingly, the clusters $C_1,\dots,C_5$ are induced by well coordinates and nearest-centroid geometry, not by pressure gradients, saturation fronts, permeability, porosity, fault distance, or other dynamic/geological attributes.
\par
For the cluster diagnostics, we report the largest evaluated grid-wide QR configuration $\Omega_{N_{\mathrm{eval,max}}}$ with $N_{\mathrm{eval,max}}=299$ selected spatial--property sensors. Because pivoted QR produces a nested ranking, $\Omega_N\subseteq\Omega_{N'}$ for $N<N'$, so $\Omega_{N_{\mathrm{eval,max}}}$ is also the union of the unique QR-selected sensors over all smaller budgets considered in the grid-wide study. Each selected spatial--property sensor $s=(c,\kappa)\in\Omega_{N_{\mathrm{eval,max}}}$ inherits the cluster label of its parent spatial cell $c\in\Omega_{\mathrm{grid}}^{xy}$, i.e., $\ell(s):=\ell(c)$. Formally, the cluster counts are
\[
n_k\;=\;\sum_{s\in\Omega_{N_{\mathrm{eval,max}}}} \mathbf{1}\!\left[\ell(s)=k\right],
\qquad
\sum_{k=1}^5 n_k\;=\;N_{\mathrm{eval,max}}=299,
\]
and the conditional probability
\[
P(C_k \mid S) \;=\; \Pr[\text{sensor located in cluster } C_k]
\]
is evaluated empirically as
\[
P(C_k\mid S)\;=\;\frac{n_k}{N_{\mathrm{eval,max}}}\;=\;\frac{n_k}{299},
\]
i.e., the fraction of the 299 QR-selected sensors that fall inside cluster \(C_k\).
\par
To define the prior \(P(C_k)\) used in the cluster analysis, we adopt the cell-fraction prior
\[
P(C_k) \;=\; \frac{|C_k|}{|\Omega_{\mathrm{grid}}^{xy}|},
\]
where \(\Omega_{\mathrm{grid}}^{xy}\) is the set of candidate grid cells.
\par
We further use the (unweighted) cell-normalized sensor density
\[
D(C_k) \;=\; \frac{n_k}{|C_k|},
\]
where \(n_k\) is the number of selected spatial--property sensors assigned to \(C_k\) and \(|C_k|\) is the number of cells in that cluster. Thus, \(D(C_k)\) is a density indicator rather than a probability and serves as the reported concentration diagnostic in the cluster analysis. The diagnostics in Table~\ref{tab:cluster_probabilities} use the largest evaluated grid-wide configuration from the $W=480$ sweep, $N_{\mathrm{eval,max}}=299$, which is a computational cutoff rather than the theoretical QR limit.
\par

\begin{table}[htbp]
  \centering
  \caption{Cluster-level diagnostics for the largest evaluated grid-wide TBMD--QR configuration.}
  \label{tab:cluster_probabilities}
  \begin{tabular}{@{}lcccc@{}}
    \toprule
    Cluster & Area fraction \(P(C_k)\) & Sensors \(n_k\) & \(P(C_k \mid S)\) & \(D(C_k)\) \\
    \midrule
    \(C1\) & 0.088 & 53  & 0.177 & 0.122 \\
    \(C2\) & 0.201 & 104 & 0.348 & 0.105 \\
    \(C3\) & 0.167 & 35  & 0.117 & 0.042 \\
    \(C4\) & 0.306 & 39  & 0.130 & 0.026 \\
    \(C5\) & 0.239 & 68  & 0.227 & 0.058 \\
    \bottomrule
\end{tabular}
\end{table}
\par
Table~\ref{tab:cluster_probabilities} shows that the grid-wide TBMD--QR ranking is non-uniform relative to the well-derived geometric partition: \(C1\) and \(C2\) are over-represented in the selected set, whereas \(C4\) is under-represented. This analysis should be interpreted as a diagnostic of QR-selected sensor concentration relative to a well-coordinate-induced geometric partition, not as direct evidence that the cluster boundaries correspond to physical flow compartments. Physical interpretation requires additional validation against dynamic or geological attributes. The observed pattern is consistent with a residual-criterion QR ranking that concentrates sensors more strongly in some partition regions than in others, but the present construction does not identify the physical cause of that concentration or establish a determinant-design property.
\par

\FloatBarrier
\FloatBarrier

The combination of \(P(C_k \mid S)\) and \(D(C_k)\) thus serves as a compact diagnostic of grid-wide TBMD--QR behavior relative to the selected partition. It highlights regions that are over- or under-represented relative to the well-derived geometric partition and provides a quantitative basis for introducing practical constraints (e.g., minimum inter-sensor spacing or quota balancing across clusters). Importantly, the method recommends measurement points rather than drilling targets; the conversion of these grid-wide recommendations to well-level priorities is handled separately in Sec.~\ref{subsec:well_significance}.

\subsection{QR-to-Well Mapping and Well Significance Analysis}
\label{subsec:well_significance}
\par
To bridge QR-guided \emph{grid-wide} sensor placement with field constraints, we quantify a post hoc overlap-frequency heuristic for individual wells by aggregating TBMD--QR selections across sensor budgets from the same grid-wide sweep with dictionary depth $W=480$ and mapping them to existing well locations. This is a post hoc diagnostic analysis, separate from the random-subset well-only reconstruction experiment of Sec.~\ref{subsec:well_based_measurements}. In this setting, a sensor is not a new well but a measurement point; any eventual deployment would still be constrained to available boreholes.
\par
Let $\mathcal{B}$ denote the set of tested sensor budgets (for example, 5, 10, 15, $\ldots$, 30). For each $N\in\mathcal{B}$, let $\Omega_N$ be the set of grid-wide TBMD--QR selected spatial--property sensors and let $\mathcal{W}$ be the set of wells with parent cells $(i_w,j_w)$. We map $\Omega_N$ to wells by retaining exact parent-cell overlaps. With a slight abuse of notation, we write $w\in\Omega_N$ if there exists at least one property index $k$ such that $(i_w,j_w,k)\in\Omega_N$. Define the raw hit count for well $w$,
\[
\tilde{S}_w \;=\; \sum_{N\in\mathcal{B}} \mathbf{1}\!\left[w \in \Omega_N\right],
\]
and its normalized significance score
\[
S_w \;=\; \frac{\tilde{S}_w - \min_{u\in\mathcal{W}}\tilde{S}_u}{\max_{u\in\mathcal{W}}\tilde{S}_u - \min_{u\in\mathcal{W}}\tilde{S}_u}
\;\in\;[0,1].
\]
(Optionally, budgets can be weighted by $\omega_N$ to emphasize larger $N$: $\tilde{S}_w=\sum_{N\in\mathcal{B}}\omega_N\,\mathbf{1}[w\in \Omega_N]$.)
Random seeds are fixed across runs for reproducibility. Here $S_w$ is an overlap-frequency heuristic only: it measures how often the parent cell of well $w$ coincides with the aggregated grid-wide QR selections across the tested budgets. Larger $S_w$ indicates stronger overlap with the QR ranking, whereas smaller $S_w$ indicates weaker overlap under this mapping. $S_w$ does not by itself prove that removing a given well would causally degrade reconstruction accuracy, nor that low-$S_w$ wells can be removed without loss. Such claims require leave-one-well-out, leave-cluster-out, or constrained reconstruction validation.
\par
\begin{figure}[htbp]
  \centering
  \includegraphics[width=0.85\textwidth, height=0.6\textheight, keepaspectratio]{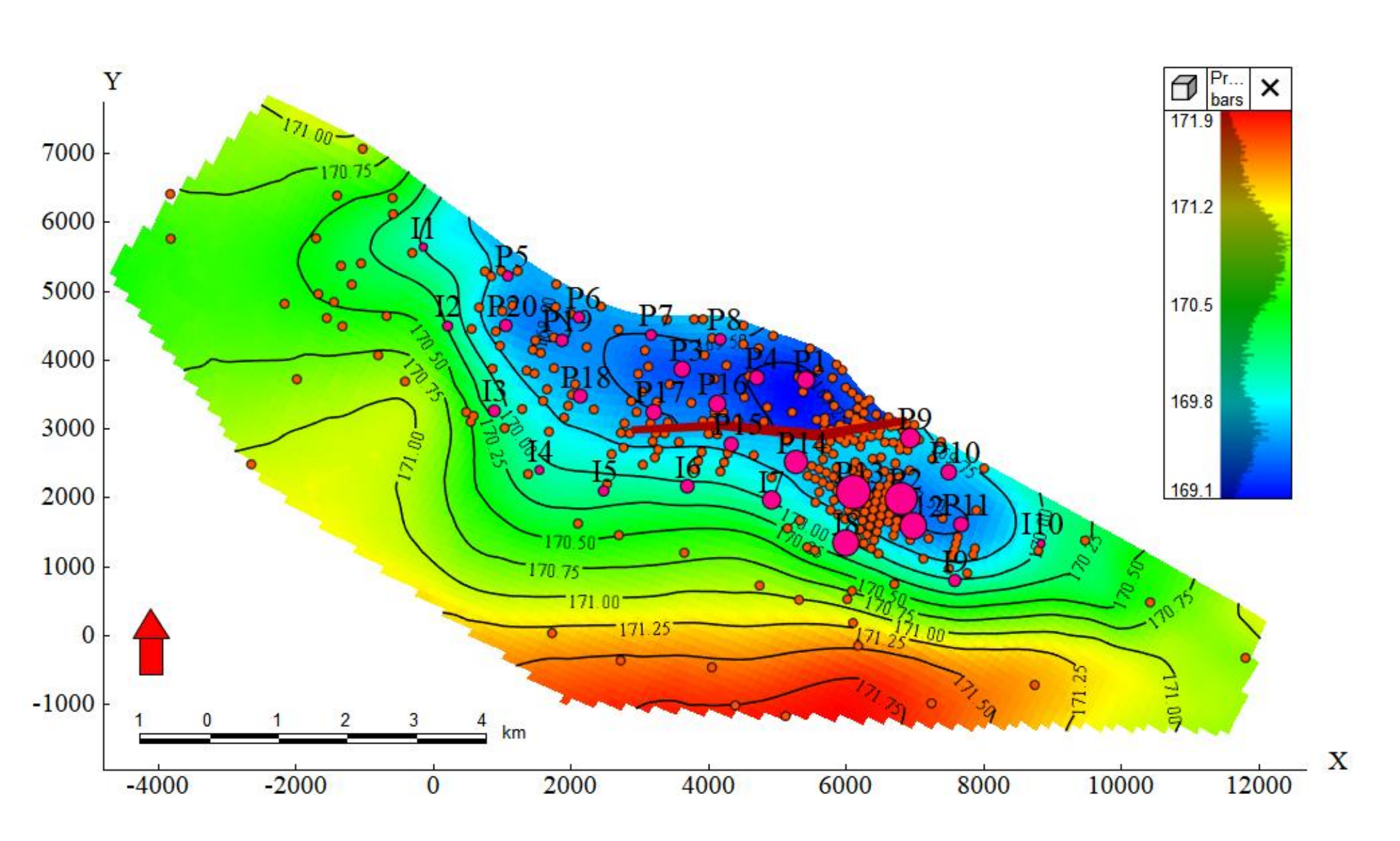}
  \caption{QR-to-well overlap-significance map. Circle radius is proportional to normalized $S_w$; red points show aggregated grid-wide QR selections.}
  \label{fig:well_informativeness_map}
\end{figure}

Figure~\ref{fig:well_informativeness_map} visualizes well significance: each magenta circle marks an existing well, with radius proportional to $S_w$; red points are the underlying grid-wide TBMD--QR sensor selections aggregated over all $N$; and the background colormap shows the normalized density of sensor hits across the grid, indicating spatial consistency of informativeness.
\par
The map reveals strong heterogeneity in overlap scores. A subset of wells achieves high $S_w$ (close to 1), indicating frequent overlap with the aggregated grid-wide QR selections, whereas other wells have consistently lower overlap under the same mapping. On Figure~\ref{fig:well_informativeness_map}, intermediate wells correspond to medium-radius magenta circles and become consistently selected only as the budget increases (e.g., $N\geq 15$). These scores summarize how frequently the existing well network coincides with the grid-wide QR ranking; they do not imply that Sec.~\ref{subsec:well_based_measurements} used QR to choose the well subsets or that the scores are ablation-validated measures of causal well criticality.
\par
Practically, these scores support a tiered prioritization hypothesis: high-$S_w$ wells are candidate wells for prioritization under a QR-overlap heuristic, mid-tier wells are secondary candidates, and low-$S_w$ wells are lower-priority candidates for further testing. Thus, the QR-to-well score should be interpreted as an empirical monitoring-priority indicator under real-world constraints rather than a causal measure of well criticality.
\par

\FloatBarrier
\FloatBarrier

\section{Comparison with Baseline Methods}
\label{sec:baseline_comparison}
The present numerical study includes two internal reference cases: full-coverage TBMD reconstruction, which checks the modal representation under complete snapshot information, and well-only random-subset reconstruction, which tests feasibility under existing infrastructure constraints. The direct grid-wide QR results should not be interpreted as a well-constrained baseline because QR is applied to the admissible grid-wide spatial--property set, whereas the well-only tests restrict measurements to the 30 existing physical wells.
\par
Comparative evidence for the core TBMD methodology was reported previously by Zhong et al.~\cite{zhong2024tbmd}, where TBMD was evaluated against POD and DMD for modal decomposition and against POD-based sparse-sensor reconstruction algorithms on cylinder-wake, airfoil-vortex, sea-surface-temperature, eigenface, and turbulent-channel-flow datasets using reconstruction-accuracy and mode/energy-loss criteria. The role of the present manuscript is different: it evaluates whether that methodological family can be extended to a coupled pressure--saturation Brugge setting with an explicit property mode, grid-wide spatial--property QR ranking, existing-well measurement operators, and QR-to-well diagnostic mapping. Therefore, the results below should be interpreted as evidence of transferability and internal consistency for this extension under the tested settings, not as a new Brugge-specific advantage claim over POD/SVD, DMD, DEIM/Q-DEIM, random sensing, or learning-based alternatives. A full task-specific benchmark on Brugge using the same train/test split, normalization, sensor budgets, admissible measurement sets, and metrics remains a useful follow-up study.

\section{Robustness and Sensitivity Analysis}
\label{sec:robustness_sensitivity}
The method demonstrated stable behavior under the tested configuration by processing 10 control realizations independently under fixed geology and aggregating held-out metrics across the resulting runs. In the well-only experiments, additional variability comes from random well subsets for each budget $N$, whereas the $N=30$ case corresponds to the fixed all-well configuration. This protocol evaluates sensitivity to well-control scenarios and random measurement subsets, but it does not constitute a geological-uncertainty ensemble.
\par
The manuscript also reports different dictionary depths for different purposes: $W=480$ for the grid-wide budget sweeps and cluster diagnostics, and a 200-mode dictionary for selected qualitative visualization. These choices should be interpreted as experiment settings rather than a full rank-sensitivity study. Broader robustness with respect to noise, rank selection, regularization, random seeds, and train/test splits requires further validation. Likewise, robustness across geological uncertainty cannot be claimed from the present fixed-geology control-realization study alone.

\section{Discussion}
\label{sec:discussion}
\par
The combined evidence from Secs.~\ref{subsec:full_coverage}--\ref{subsec:well_significance} indicates that the proposed TBMD--QR--TBCS extension is internally consistent across three complementary levels. Section~\ref{subsec:full_coverage} establishes the full-coverage modal-representation check and the direct grid-wide QR-placement results. Section~\ref{subsec:well_based_measurements} then evaluates reconstruction quality under existing-well constraints via random well subsets. Finally, Secs.~\ref{subsec:probabilistic_analysis}--\ref{subsec:well_significance} analyze the grid-wide QR ranking and map it onto the existing well infrastructure, thereby characterizing how the QR ranking is distributed relative to the induced partition and the existing well network.
\par
The present results evaluate the transferability of the TBMD methodology to coupled pressure--saturation reservoir fields and measurement operators reflecting grid-wide and existing-well sensing, demonstrating its extension behavior under fixed-geology control realizations.
\par
It is important to distinguish between the representative grid-wide reconstruction maps, which use a 200-mode dictionary only for qualitative visualization, and the full grid-wide budget sweep plus clustering analysis, which use dictionary depth $W=480$. Under this latter configuration, the largest evaluated budget is $N_{\mathrm{eval,max}}=299$, chosen as a practical upper cutoff for the reported sweep rather than as the mathematical QR limit. Therefore, $N_{\mathrm{eval,max}}=299 < W=480$, fully consistent with the admissibility condition $N\leq\min(W,|\Omega_{\mathrm{grid}}|)$.
\par
A key quantitative finding is the early gain with sensor budget. In the single-property well-only random-subset case (Sec.~\ref{subsubsec:well_based_single}), increasing sensors from \(N=1\) to \(N=10\) reduces relative error from about \(0.62\) to \(0.17\), increases SSIM from about \(0.47\) to \(0.90\), and raises PSNR from about \(27\) to \(34.5\,\mathrm{dB}\). In the joint multi-property case (Sec.~\ref{subsubsec:well_based_multi}), the same increase (\(N=1\to10\)) reduces relative error from about \(0.57\) to \(0.20\), increases SSIM from about \(0.47\) to \(0.88\), and raises PSNR from about \(33.8\) to \(37\,\mathrm{dB}\). In the reported random-subset protocol, most observed gains occur within the first \(10\text{--}15\) wells, while additional wells mainly improve empirical stability in the reported curves. Because each \(N\) is based on random well subsets, these curves quantify feasibility under the existing-well constraint rather than QR-ranked selection among wells.
\par
The spatial statistics of the grid-wide QR ranking show non-uniform concentration relative to the well-derived geometric partition. From Table~\ref{tab:cluster_probabilities}, sensor allocation is strongly non-uniform: \(P(C_1\mid S)=0.177\) vs. \(P(C_1)=0.088\) and \(P(C_2\mid S)=0.348\) vs. \(P(C_2)=0.201\), while \(C_4\) is under-represented (\(0.130\) vs. \(0.306\)). The cell-normalized sensor density is likewise highest in \(C_1\) and \(C_2\) (\(0.122\), \(0.105\)) and lowest in \(C_4\) (\(0.026\)). This pattern is consistent with the hypothesis that the QR ranking emphasizes some regions of the induced partition more than others. However, because \(C_1,\dots,C_5\) are induced by well coordinates rather than pressure gradients, saturation fronts, permeability, porosity, fault distance, or other dynamic/geological attributes, the cluster analysis should be interpreted only as a partition-relative diagnostic, not as direct evidence of physical flow compartments.
\par
From an operational viewpoint, the QR-to-well score quantifies how frequently existing wells coincide with QR-selected informative spatial locations. It should be interpreted as an empirical monitoring-priority indicator rather than a causal measure of well criticality: causal claims about degradation from removing a well or the safety of deprioritizing a low-\(S_w\) well require leave-one-well-out, leave-cluster-out, or other constrained reconstruction validation. A practical caveat is that the Sec.~\ref{subsec:well_based_measurements} metric curves are not strictly monotone because each \(N\) is evaluated on random well subsets.

\FloatBarrier

\section{Limitations}
\label{sec:limitations}
A limitation of the present study is the absence of a direct benchmark against POD, DEIM, random sensing, or learning-based alternatives on the specific Brugge configuration. A task-specific comparison with identical train/test splits, sensor budgets, and metrics remains necessary to establish explicit performance advantages.
\par
The present experiments also use fixed geology and vary only well-control conditions, so they do not quantify robustness to geological uncertainty. The well-only experiments are random-subset feasibility tests rather than QR-ranked well-selection studies. The QR-to-well scores are overlap-frequency diagnostics and do not establish causal well criticality. Robustness with respect to measurement noise, dictionary rank, ADMM regularization, random seeds, train/test splits, and runtime/hardware variability remains to be quantified.

\section{Conclusion and Outlook}
\label{sec:conclusion}
\par
This paper extends a tensor-based modal decomposition (TBMD) sparse reconstruction framework to coupled pressure--saturation reservoir simulation data. By adding a fourth-order tensor formulation with an explicit property mode, the framework retains the gridded spatial organization and cross-property structure of the simulated fields. We introduced a grid-wide spatial--property sensor ranking based on tensor QR pivoting and evaluated snapshot-wise full-field reconstructions under practical existing-well measurement constraints. 

On the Brugge benchmark, consisting of 10 control realizations with fixed geology, the reconstruction metrics systematically improved with the number of instrumented wells. Specifically, increasing the well count from 1 to 10 reduced the relative error from 0.57 to 0.20 and raised the structural similarity from 0.47 to 0.88. The proposed QR-to-well mapping produced non-uniform, partition-relative diagnostics that prioritize informative existing wells. Overall, the 4D TBMD extension provides an interpretable and compact computational workflow for evaluating sparse monitoring feasibility under realistic well-only constraints. 

Future work will investigate robustness to geological uncertainty, integrate static properties into the tensor dictionary, and benchmark computational runtime against alternative reduced-order and learning-based monitoring approaches.

\section*{Data and Code Availability}

\noindent\textbf{Data availability.}
The study uses the Brugge reservoir simulation benchmark described in the cited benchmark references. The processed pressure and oil-saturation tensors supporting the findings of this study are available from the corresponding author upon reasonable request.

\par
\begingroup\sloppy
\noindent\textbf{Code availability.}
The source code and configuration files used in this study are publicly available at \url{https://github.com/denis-samatov/tensor_based_modal_decomposition_method}. A persistent Zenodo archive will be published alongside the final manuscript. 
\par\endgroup

\section*{Declaration of Competing Interest}

The authors declare that they have no known competing financial interests or personal relationships that could have appeared to influence the work reported in this paper.

\section*{CRediT Author Statement}

D. Samatov: Conceptualization, Methodology, Software, Validation, Formal analysis, Investigation, Visualization, Writing -- original draft.

\par
B. Merzlikin: Supervision, Methodology, Writing -- review \& editing.

\par
G. Shishaev: Data curation, Validation, Writing -- review \& editing.

\section*{Funding}

This research did not receive any specific grant from funding agencies in the public, commercial, or not-for-profit sectors.

\section*{Acknowledgements}

The authors acknowledge the Brugge benchmark contributors and the cited benchmark documentation for providing a widely used reservoir-simulation benchmark for computational studies.

\section*{Declaration of Generative AI and AI-assisted Technologies}

The authors declare that no generative AI or AI-assisted technologies were used in the writing or scientific development of this manuscript.

\bibliographystyle{elsarticle-num}
\bibliography{references}

\end{document}